\documentclass[sigconf, 9pt]{acmart}

\AtBeginDocument{%
  }

\usepackage{amsmath}
\usepackage{amsfonts}
\usepackage{bm}
\usepackage{multirow}
\usepackage{tabularx}
\usepackage{array}
\usepackage{color}
\usepackage{enumitem}
\usepackage{makecell}
\usepackage{xcolor}
\usepackage{colortbl}
\usepackage{booktabs}
\usepackage{algorithm}
\usepackage{algpseudocode}
\usepackage{pifont}
\usepackage{bbding}

\definecolor{my_green}{RGB}{227, 254, 223}
\definecolor{my_blue}{RGB}{237, 238, 254}

\newcolumntype{C}{>{\centering\arraybackslash}X}
\newcolumntype{B}{>{\columncolor{my_blue}}C}
\newcommand{\proj}{\textsl{SelfReplay}}

\newcommand{\rev}[1]{{\leavevmode\color{black}{#1}}}
\newcommand{\shepherd}[1]{{\leavevmode\color{black}{#1}}}
\newcommand{\shepherdrev}[1]{{\leavevmode\color{black}{#1}}}

\newcommand{\eg}{\textit{e.g.},}

\definecolor{LightCyan}{rgb}{0.88,1,1}

\begin{document}

\acmYear{2025}\copyrightyear{2025}
\setcopyright{rightsretained}
\acmConference[SenSys '25]{The 23rd ACM Conference on Embedded Networked Sensor Systems}{May 6--9, 2025}{Irvine, CA, USA}
\acmBooktitle{The 23rd ACM Conference on Embedded Networked Sensor Systems (SenSys '25), May 6--9, 2025, Irvine, CA, USA}
\acmDOI{10.1145/3715014.3722066}
\acmISBN{979-8-4007-1479-5/25/05}

\title[\proj{}: Adapting Self-Supervised Sensory Models via Adaptive Meta-Task Replay]{\texorpdfstring{\proj{}: Adapting Self-Supervised \\Sensory Models via Adaptive Meta-Task Replay}{\proj{}: Adapting Self-Supervised Sensory Models via Adaptive Meta-Task Replay}}


\author{Hyungjun Yoon}
\affiliation{
  \institution{KAIST}
  \city{}
  \country{}
}
\email{hyungjun.yoon@kaist.ac.kr}

\author{Jaehyun Kwak}
\affiliation{
  \institution{KAIST}
  \city{}
  \country{}
}
\email{jaehyun98@kaist.ac.kr}

\author{Biniyam Aschalew Tolera}
\affiliation{
  \institution{KAIST}
  \city{}
  \country{}
}
\email{binasc@kaist.ac.kr}

\author{Gaole Dai}
\affiliation{
  \institution{Nanyang Technological University}
  \city{}
  \country{}
}
\email{gaole001@e.ntu.edu.sg}

\author{Mo Li}
\affiliation{
  \institution{HKUST}
  \city{}
  \country{}
}
\email{lim@cse.ust.hk}

\author{Taesik Gong}
\affiliation{
  \institution{UNIST}
  \city{}
  \country{}
}
\email{taesik.gong@unist.ac.kr}

\author{Kimin Lee}
\affiliation{
  \institution{KAIST}
  \city{}
  \country{}
}
\email{kiminlee@kaist.ac.kr}

\author{Sung-Ju Lee}
\affiliation{
  \institution{KAIST}
  \city{}
  \country{}
}
\email{profsj@kaist.ac.kr}

\renewcommand{\shortauthors}{Yoon et al.}

\ccsdesc[500]{Human-centered computing~Ubiquitous and mobile computing systems and tools}
\ccsdesc[500]{Computing methodologies~Machine learning}

\keywords{Self-Supervised learning, Mobile Sensing, Domain Adaptation}

\begin{abstract}

\rev{Self-supervised learning enables effective model pre-training on large-scale unlabeled data, which is crucial for user-specific fine-tuning in mobile sensing applications. However, pre-trained models often face significant domain shifts during fine-tuning due to user diversity, leading to performance degradation. To address this, we propose \proj{}, an adaptive approach designed to align self-supervised models to different domains. \proj{} consists of two stages: MetaSSL, which leverages meta-learning with self-supervised learning to pre-train domain-adaptive weights, and ReplaySSL, which further adapts the pre-trained model to each user’s domain by replaying the meta-learned self-supervised task with a few user-specific samples. This produces a personalized model tailored to each user. Evaluations on mobile sensing benchmarks demonstrate that \proj{} outperforms existing baselines, improving the F1-score by 9.4\%p on average. \shepherd{On-device analyses on a commodity smartphone show the efficiency of \proj{}'s adaptation step, required just once after deployment, with SimCLR completing in only 10 seconds while using less than 100MB of memory.}}

\end{abstract}

\maketitle

\sloppy
\section{Introduction}
\label{sec:introduction}

\rev{Large-scale pre-training has become fundamental in developing models that generalize across diverse applications. Self-supervised learning~\cite{jing2020self}, in particular, enables models to leverage vast amounts of unlabeled data, making it a powerful approach for building foundation models. In mobile sensing applications such as contactless authentication~\cite{wang2021ear, liu2022recent}, sign language translation~\cite{hou2019signspeaker, park2021enabling}, and mobile health monitoring~\cite{zhang2019pdmove, song2020spirosonic}, labeled data is often scarce and costly to acquire, which makes self-supervised methods especially valuable. Self-supervised techniques such as predictive coding~\cite{haresamudram2021contrastive, haresamudram2023investigating}, contrastive learning~\cite{tang2020exploring}, and multi-task learning~\cite{saeed2019multi} have demonstrated effectiveness in sensory applications by reducing reliance on labeled data in pre-training.

However, challenges arise once pre-trained models are deployed for fine-tuning across different environments. In mobile sensing, data collected in diverse environments varies widely due to differences in users and device settings (\textit{e.g.}, sensor placement and sampling rate)~\cite{ustev2013user}. These variations introduce a \textit{domain shift}, where models trained in one domain underperform when applied to others~\cite{stisen2015smart}. To highlight this challenge in the self-supervised setting, we conducted an empirical analysis using a self-supervised pre-training method~\cite{haresamudram2023investigating} for human activity recognition~\cite{gong2019metasense} (details in Section~\ref{sec:domain_shift_effect}). Figure~\ref{fig:intro_motivation} compares the performance of a pre-trained model when fine-tuned within the same source domain versus on a different target domain. The results show that performance declines substantially when fine-tuning is conducted on a different domain. This underscores the challenge of deploying self-supervised models across diverse mobile sensing environments.

\begin{figure}[t]
    \centering
    \includegraphics[width=0.95\columnwidth]{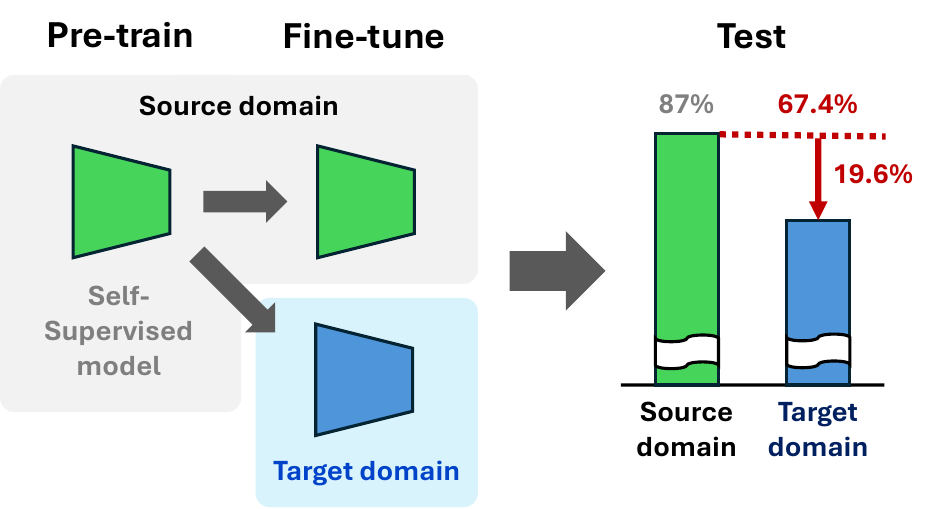}
    \vspace{-3pt}
    \caption{\rev{Illustration of domain shift on a self-supervised model pre-trained in one domain and fine-tuned in another for human activity recognition. Fine-tuning on the target domain results in a 19.6\% F1-score drop (87\% vs. 67.4\%).}
    \Description{Illustration of domain shift in a pre-trained model trained using self-supervised learning and fine-tuned for a human activity recognition task.}}
    \label{fig:intro_motivation}
\end{figure}

Training a domain-specific model using target domain data is straightforward but infeasible due to the cost of gathering sufficient data from each user. Existing research includes domain generalization~\cite{qian2021latent, qin2022domain, lu2022semantic} that trains models with domain-invariant features, and domain adaptation~\cite{chang2020systematic, gong2019metasense, zhou2020xhar, wang2018stratified} that leverages a small amount of target domain data to achieve domain-specific performance. However, they primarily focus on supervised learning scenarios, making applying to domain shifts arising from self-supervised pre-training difficult.

To address the problem, we propose \proj{}, an \textit{adaptive meta-task replay} approach for adapting self-supervised models to different domains. Figure~\ref{fig:intro_overview} illustrates \proj{} alongside a standard pre-training and fine-tuning setting. We design \proj{} with two key components: (i) MetaSSL generates domain-adaptive pre-trained models and (ii) ReplaySSL adapts the pre-trained model to the target domain. MetaSSL leverages meta-learning over self-supervised objectives, structuring pre-training into multiple few-shot tasks by domain. This enables models to ``learn to self-supervise'' on only a few domain-specific data, following meta-learning’s concept of ``learning to learn.'' After pre-training, ReplaySSL adapts the model to the target domain by replaying the meta-learned self-supervised task with a few samples, creating a \textit{personalized} model. The adapted model is then fine-tuned for the final application task.

We evaluate \proj{} on mobile sensing datasets~\cite{gong2019metasense, stisen2015smart, reiss2012introducing, altun2010comparative} by simulating domain shifts across different users and devices. Our experiments show that when self-supervised models are fine-tuned on heterogeneous domains, \proj{} consistently outperforms domain generalization and adaptation baselines~\cite{zhang2022towards, lee2023self}, achieving an average F1-score improvement of 9.4\%p. Importantly, \proj{} is agnostic to self-supervised learning objectives, making it compatible with various approaches. We apply \proj{} to contrastive learning~\cite{tang2020exploring}, predictive coding~\cite{haresamudram2023investigating}, and multi-task learning~\cite{saeed2019multi}, demonstrating its effectiveness across different self-supervised methods. To assess practical feasibility, we also measure \proj{}’s computational overhead on three edge devices. On a standard smartphone, ReplaySSL with SimCLR completes in just 10 seconds while using less than 100MB of memory. As the adaptation step runs only once after obtaining the pre-trained model, it introduces minimal computational load for the user.

\begin{figure}[t]
    \centering
    \includegraphics[width=0.95\columnwidth]{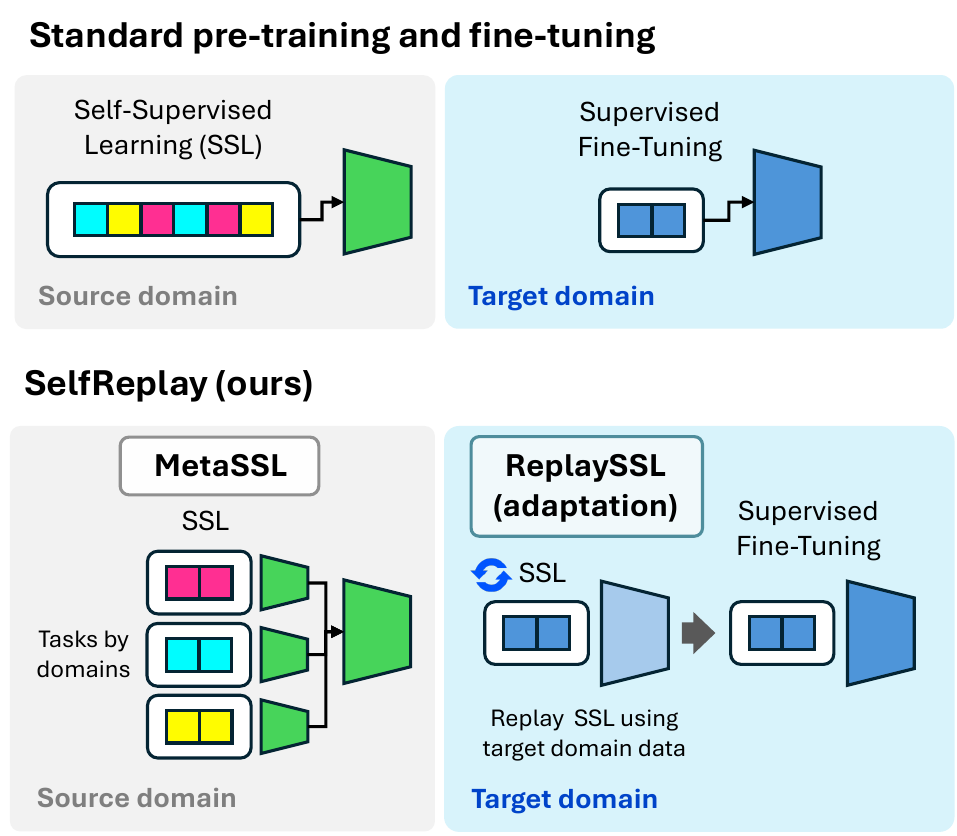}
    \vspace{-3pt}
    \caption{\rev{A comparison between the standard pre-training and fine-tuning (top) and \proj{} (bottom). Components in the grey box are performed in the source domain, while those in the blue box are performed in the target domain.}}
    \label{fig:intro_overview}
    \Description{A comparison between the standard pre-training and fine-tuning (top) and \proj{} (bottom).}
\end{figure}

We summarize our main contributions as follows:

\begin{itemize}[left=1em, topsep=0pt, partopsep=0pt]

\item We investigate the domain shift issue when diverse users deploy and fine-tune self-supervised models. We reveal that the domain shift leads to performance degradation.

\item We propose \proj{}, a method for adapting self-supervised models to different domains via adaptive meta-task replay.

\item We perform evaluations using mobile sensing datasets, showing that \proj{} outperforms domain generalization and adaptation baselines, achieving an average F1-score improvement of 9.4\%p.

\item We assess the computational overhead of deploying \proj{} on mobile devices, demonstrating that the adaptation step completes in 10 seconds with less than 100MB of memory usage for SimCLR on an off-the-shelf smartphone.
\end{itemize}
}
\section{Related Work}

\begin{figure*}
    \centering
    \includegraphics[width=0.95\textwidth]{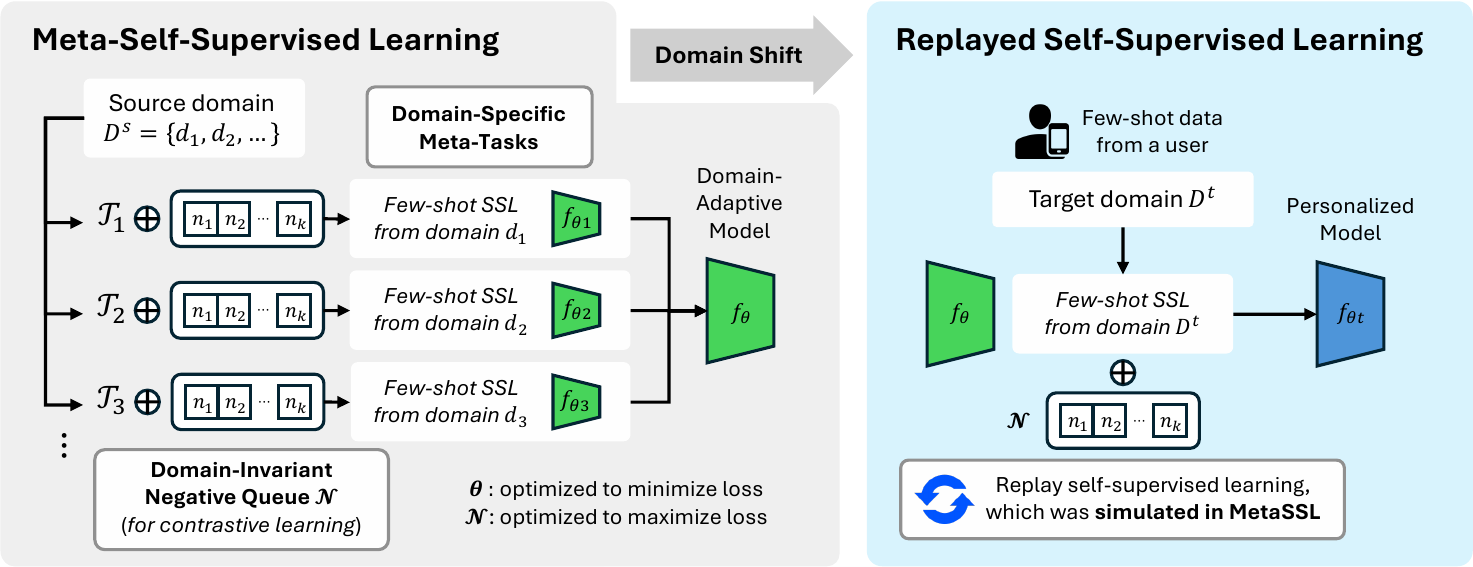}
    \vspace{-3pt}
    \caption{\rev{Overview of \proj{}. The model is first pre-trained through MetaSSL in the source domain to develop domain adaptability, followed by ReplaySSL to adapt the model to the target domain.}}
    \Description{Overview of \proj{}.}
    \label{fig:method_overview}
    \vspace{-5pt}
\end{figure*}

\subsection{Self-Supervised Learning}

Self-supervised learning trains models using an auxiliary task that can be defined without labels, which enables learning generalized features of the data. Among numerous approaches, we focus on the methods applied for mobile sensing~\cite{haresamudram2022assessing}. multi-task Learning~\cite{saeed2019multi} utilized multiple types of synthetic augmentations on the data and trained task prediction networks to infer the occurrence of the augmentation. Sensor-specific augmentations were selected to make the model learn sensory properties. Recent work focuses on using contrastive learning~\cite{jaiswal2020survey}, which generates augmented views of data and trains the model intending to maximize the similarity between the augmented views. Existing methods such as MoCo~\cite{he2020momentum} and SimCLR~\cite{chen2020simple} were applied to mobile sensing~\cite{wang2022sensor, tang2020exploring} by using sensory augmentations to generate views. The temporal property of time-series data is utilized for generating views in a recent work~\cite{eldele2023self}. Taking into account the multi-modality, Cosmo~\cite{ouyang2022cosmo}, COCOA~\cite{deldari2022cocoa}, and ColloSSL~\cite{jain2022collossl} utilized contrastive learning to maximize the similarity between the embeddings driven from different modalities in the same context. Contrastive predictive coding (CPC)~\cite{haresamudram2021contrastive, haresamudram2023investigating} defined another type of task, predicting the embedding of future segments within the data based on the previously aggregated embeddings. In a similar context, masked-reconstruction-based methods~\cite{haresamudram2020masked, xu2021limu} have been explored for mobile sensing, using the task of reconstructing the synthetically masked segment within the data.

While self-supervised models are known to be generalizable across diverse tasks, the potential performance decline when applied to different domains~(shown in Section~\ref{sec:domain_shift_effect}) is overlooked. Our work differs from the prior research in exploring the domain shift problem between self-supervised pre-training and fine-tuning.

\subsection{Domain Generalization and Domain Adaptation}

\shepherd{Domain generalization (DG)~\cite{wang2022generalizing} mitigates domain shifts by learning domain-invariant features via adjusted objectives~\cite{muandet2013domain, li2019episodic}, adversarial discrimination~\cite{li2018domain}, and domain-information minimization~\cite{wang2020heterogeneous, zhou2021domain}. Recent work incorporated meta-learning~\cite{li2018learning, qiao2020learning} and self-supervised learning~\cite{kim2021selfreg, yao2022pcl} to define domain-invariant training objectives. In mobile sensing, GILE~\cite{qian2021latent} disentangled domain-specific information, while SDMix~\cite{lu2022semantic} employed semantic-aware augmentations to achieve DG tailored to activity recognition. For placement shifts, position- and orientation-agnostic methods~\cite{user_device_independent, s17081838} rely on data preprocessing and feature transformations but remain limited to a single domain factor and are less adaptable to user differences or other domain variations. Domain adaptation (DA)~\cite{wilson2020survey} is more suitable for our scenario, as it allows fine-tuning with user-collected data. Common approaches rely on unlabeled or limited-labeled target data~\cite{ganin2015unsupervised, rahman2020correlation} and employ strategies such as feature matching, confusion maximization~\cite{chang2020systematic}, and transfer learning for activity recognition~\cite{wang2018stratified, lu2021cross, khan2018scaling}. MetaSense~\cite{gong2019metasense} introduced a meta-learning approach followed by few-shot adaptation to create domain-specific models. DAPPER~\cite{gong2023dapper} is another line of research for estimating the expected performance of DA in mobile sensing.

However, these approaches assume the availability of source-domain labels and thus are incompatible with our unsupervised pre-training. DARLING~\cite{zhang2022towards} addresses domain shift without source labels by integrating conditional optimization that modifies the contrastive loss per domain. Likewise, ContrastSense~\cite{contrastsense} targets unsupervised pre-training by introducing a contrastive loss that minimizes inter-domain discrepancies. 
While these methods effectively generalize across domains, our approach differs by training \textit{domain-adaptive} pre-trained models (MetaSSL), and then leveraging available target-domain data (\textit{i.e.}, fine-tuning data) to train domain-specific models, thereby achieving superior performance through an additional adaptation step (ReplaySSL).}

\subsection{Unsupervised Meta-Learning}

We consider unsupervised meta-learning (UML)~\cite{khodadadeh2019unsupervised, khodadadeh2020unsupervised, lee2023self} methods due to their effectiveness in few-shot adaptation, which is also applicable to our unsupervised pre-training scenario. Traditional methods~ employ pseudo-labeling data through augmentation~\cite{khodadadeh2019unsupervised} or generative methods~\cite{khodadadeh2020unsupervised}, followed by supervised meta-learning~\cite{andrychowicz2016learning} using the generated labels. Set-SimCLR~\cite{lee2023self}, during pre-training, trains a set encoder by creating sets of augmented samples from the same data, employing contrastive learning to maximize agreement between set embeddings. In fine-tuning, it composes sets of data by classes, generating class prototypes using the set encoder to initialize the classifier's parameters. These prototypes enable rapid adaptation for further few-shot fine-tuning. However, our approach differs in that we perform the adaptation to refine the encoder for the target domain, while Set-SimCLR primarily focuses on making the following classifier adaptable to few-shot fine-tuning. Our evaluation (Section~\ref{sec:overall_results}) demonstrates the superior performance of our approach in mobile sensing scenarios.
\section{Method}

\rev{We present \proj{}, an approach for adapting self-supervised models to diverse domains via adaptive meta-task replay. In Section~\ref{sec:problem_formulation}, we formulate the domain shift problem that arises when self-supervised models are deployed to different target domains. Our solution, illustrated in Figure~\ref{fig:method_overview}, incorporates two key strategies: (i) Meta-Self-Supervised Learning (MetaSSL) to pre-train domain-adaptive weights, and (ii) Replayed Self-Supervised Learning (ReplaySSL) to adapt pre-trained models to the target domain with only few samples. We detail the design of MetaSSL in Section~\ref{sec:metassl} and ReplaySSL in Section~\ref{sec:replayssl}.}

\subsection{Problem Formulation} \label{sec:problem_formulation}

\noindent \textbf{Domain Shift from Pre-Training.} \shepherdrev{Pre-training generates representation models using large-scale data, but obtaining corresponding labels is often expensive or restricted. For example, in activity recognition, asking users to report their activity every hour is costly. In health applications, labels often contain sensitive personal information (\textit{e.g.}, medical diagnoses and mental health assessments), sometimes making label acquisition infeasible. To address this, models are pre-trained on unlabeled data before being adapted to specific tasks. Existing large-scale efforts have produced models trained solely on unlabeled datasets, such as Google's 40 million hours of physiological data~\cite{narayanswamy2025scaling} and UK-Biobank's 700,000 person-days of accelerometer data~\cite{yuan2024self}.

We target scenarios where users utilize the pre-trained models for their own applications. Pre-trained models provide rich feature representations, enabling users to train models with only a small amount of data. For example, when setting up a fitness app, a smartphone might prompt the user to stay, walk, or run for 30 seconds, allowing the model to fine-tune running detection with just a few labeled samples. Similarly, for gesture recognition, leveraging pre-trained models trained on UK-Biobank, a user might define and repeat a custom hand gesture a few times for fine-tuning.

To summarize, our scenario consists of three steps: (i) a model is pre-trained on a large amount of unlabeled data, (ii) it is fine-tuned with only a few labeled samples, and (iii) it is evaluated on test data. Pre-training occurs on a source domain $D^s$, which is distinct from the target domain $D^t$ used for fine-tuning and testing.}



\noindent \textbf{Availability of Domain Labels.} During pre-training, we assume access to domain labels within $D^s$. Domain labels (\textit{e.g.}, device type, or anonymized user identifiers) are generally available as metadata, allowing us to distinguish data from different sources without exposing sensitive information.

\noindent \textbf{Few-Shot Fine-Tuning.} After deployment, users fine-tune the pre-trained model using their own data from the target domain $D^t$. \shepherd{We target scenarios where users can easily collect a small amount of labeled data, such as by repeating a few representative actions per class, to customize the model for their specific needs. Therefore, we assume that only a few labeled samples (\textit{e.g.}, 10 per class) are available for fine-tuning.}

\subsection{Meta-Self-Supervised Learning}
\label{sec:metassl}

\rev{We propose an approach to adapt self-supervised models to a target domain, even when only a few data samples are available. To enable this, we enhance self-supervised pre-training to produce domain-adaptive weights, allowing the model to align with a specific domain using minimal data.

Our approach, \textit{Meta-Self-Supervised Learning (MetaSSL)}, prepares models for adaptation by combining meta-learning with self-supervised objectives. Meta-learning~\cite{andrychowicz2016learning}, often referred to as ``learning to learn,'' trains models to be fine-tuned effectively in new conditions with a few data. Inspired by its efficacy in traditional supervised settings, we design MetaSSL as a method for ``learning to self-supervise.'' MetaSSL produces a model that is adaptive to few-shot self-supervised learning. To further adapt the model, we introduce an additional step that replays self-supervised training using data from the target domain (Section~\ref{sec:replayssl}).

Our implementation of MetaSSL builds on Model-Agnostic Meta-Learning (MAML)~\cite{finn2017model}, which optimizes a model’s initial weights for further updates with minimal data, making it well-suited to our approach. Additionally, MAML’s flexibility allows MetaSSL to be applied across various self-supervised learning objectives~\cite{oord2018representation, chen2020simple, saeed2019multi} (Section~\ref{sec:ssl_results}). We detail the MetaSSL procedure in the next sections.

\begin{algorithm}[t]
\caption{Domain-Specific Task Generation}
\label{alg:task_generation}
\begin{flushleft}
\textbf{Inputs:} Pre-train dataset $X$, number of tasks $M$, and number of domain-specific tasks $M^{\tt dom}<M$
\end{flushleft}
\begin{algorithmic}[1]
\State Initialize empty task set $\mathcal{T}$
\For{$i \in \{1, 2, ..., M\}$} 
    \If{$i \leq M^{\tt dom}$} 
        \State Select domain $d_i$ from $D^s$ uniformly at random
        \State Select $K$ samples with domain $d_i$ randomly:
        \Statex \hspace{\algorithmicindent}\hspace{\algorithmicindent} $\mathcal{S}_i=\{x \in X | \operatorname{dom}(x) = d_i \}$ such that $|\mathcal{S}_i| = K$
        \State Select another set of $K$ samples with $d_i$ randomly:
        \Statex \hspace{\algorithmicindent}\hspace{\algorithmicindent} $\mathcal{Q}_i=\{x \in X | \operatorname{dom}(x) = d_i \}$ such that $|\mathcal{Q}_i| = K$
    \Else
        \State Select $K$ samples randomly: 
        \Statex\hspace{\algorithmicindent}\hspace{\algorithmicindent} $\mathcal{S}_i=\{x \in X\}$ such that $|\mathcal{S}_i| = K$
        \State Select another set of $K$ samples randomly: 
        \Statex\hspace{\algorithmicindent}\hspace{\algorithmicindent} $\mathcal{Q}_i=\{x \in X\}$ such that $|\mathcal{Q}_i| = K$
    \EndIf
    \State $\mathcal{T}_i \leftarrow (\mathcal{S}_i,\mathcal{Q}_i)$
    \State Update a set of task $\mathcal{T} \leftarrow \mathcal{T} \cup \mathcal{T}_i$
\EndFor \\
\Return Task set $\mathcal{T}$
\end{algorithmic}
\end{algorithm}

\noindent \textbf{Domain-Specific Task Generation.} MetaSSL simulates few-shot self-supervised learning through a set of \textit{tasks}, $\mathcal{T}$, that emulate training and testing on a limited data. By default, each task $\mathcal{T}_i$ involves optimizing \textit{domain-specific weights} $\theta_i$ on a small dataset, the \textit{support set} $\mathcal{S}_i$. The optimized weights are then evaluated on a separate \textit{query set} $\mathcal{Q}_i$. The evaluation results from all tasks are combined to compute a single loss, which guides the optimization of the model’s \textit{global weights} $\theta$ and reinforces its ability to adapt effectively across tasks.

Each task $\mathcal{T}_i$ is configured to operate within a single domain $d_i$, selected randomly from the source domain $D^s$ using domain labels. This setup enables each task to simulate few-shot self-supervised learning and test within a specific domain, thus facilitating domain-specific weight optimization. We refer to these as \textit{domain-specific tasks}. Following prior work~\cite{gong2021adapting}, we introduce a small proportion (\textit{e.g.}, 30\%) of \textit{multi-conditioned tasks}---tasks generated from random, domain-agnostic samples---to reduce the risk of overfitting to particular domains. These multi-conditioned tasks, formed from mixed-domain samples, act as synthetic domains that add diversity to the training. Algorithm~\ref{alg:task_generation} summarizes our domain-specific task generation process.

\begin{algorithm}[t]
\caption{Meta-Self-Supervised Learning (MetaSSL)}
\label{alg:domain_adaptive_ssl}
\begin{flushleft}
\textbf{Inputs:} Pre-train dataset $X$, number of tasks $M$, number of domain-specific tasks $M^{\tt dom}$, model weights $\theta$, and self-supervised loss function $\mathcal{L}_\texttt{SSL}$
\end{flushleft}
\begin{algorithmic}[1]
\For{epochs}
    \State $\mathcal{T} =$ TaskGeneration($X, M, M^{\tt dom}$) \Comment{Algorithm~(\ref{alg:task_generation})}
    \For{$\mathcal{T}_i =(\mathcal{S}_i, \mathcal{Q}_i) \in \mathcal{T}$}
        \State Optimize task-specific weights using $\mathcal{S}_i$:
        \Statex\hspace{\algorithmicindent}\hspace{\algorithmicindent} $\theta_i \gets \theta-\alpha \nabla_\theta \mathcal{L}_\texttt{SSL}(\theta; \mathcal{S}_i)$ 
        \State Evaluate loss of updated model on $\mathcal{Q}_i$:
        \Statex\hspace{\algorithmicindent}\hspace{\algorithmicindent} Compute $\mathcal{L}_\texttt{SSL}(\theta_i; \mathcal{Q}_i)$
    \EndFor
    \State Update weights $\theta \gets \theta-\beta\nabla _\theta \sum_{i} \mathcal{L}_\texttt{SSL}(\theta_i; \mathcal{Q}_i)$
\EndFor
\end{algorithmic}
\end{algorithm}

\noindent \textbf{MetaSSL Optimization.} MetaSSL pre-trains domain-adaptive weights by simulating few-shot self-supervised learning tasks within domains. Each domain-specific task $\mathcal{T}_i$, generated from a single domain, trains domain-specific weights $\theta_i$ on a support set $\mathcal{S}_i$ and evaluates performance on a corresponding query set $\mathcal{Q}_i$. Both training and evaluation use a self-supervised loss function $\mathcal{L}_\texttt{SSL}$ to optimize task-specific weights to the domain. For example, when contrastive learning~\cite{chen2020simple} is used, the loss function applied to $\mathcal{S}_i$ in a domain-specific task (same for $\mathcal{Q}_i$) is:
\begin{align}
\notag &\mathcal{L}_\texttt{SSL} (\theta_i; \mathcal{S}_i = \{ x \mid \operatorname{dom}(x) = d_i \}) =\\
&- \sum_{x_j\in \mathcal{S}_i} \log\left(\frac{e^{\text{sim}(f_{\theta_i}(x'_j), f_{\theta_i}(x''_j))}}{e^{\text{sim}(f_{\theta_i}(x'_j), f_{\theta_i}(x''_j))} + \sum_{k\neq j} e^{\text{sim}(f_{\theta_i}(x'_j), f_{\theta_i}(x_k))}}\right),
\label{eq:method_sslloss}
\end{align}
\noindent where the parameterized encoder $f_{\theta_i}$ is optimized to maximize the similarity between two augmented views of the same data point ($x^{\prime}_i$ and $x^{\prime \prime}_i$) while minimizing similarity with other samples $x_k$, \textit{i.e.}, negative samples. Here, $\mathcal{S}_i$ contains samples from a specific domain $d_i$ when it is from a domain-specific task, ensuring that the contrastive loss is computed within a single domain. Note that our framework is agnostic to specific self-supervised methods, allowing $\mathcal{L}_\texttt{SSL}$ to be replaced with various objectives (\textit{e.g.}, contrastive predictive coding~\cite{oord2018representation} and multi-task learning~\cite{saeed2019multi}).

After evaluating the domain-specific weights on each $\mathcal{Q}_i$, results are aggregated across tasks to optimize the global model weights $\theta$. This iterative process results in global model weights that are highly adaptable for few-shot self-supervised learning across different domains. Algorithm~\ref{alg:domain_adaptive_ssl} outlines the full procedure of MetaSSL. $\alpha$ and $\beta$ denote the learning rates for domain-specific training of $\theta_i$ and global model weights $\theta$ update.

\noindent \textbf{Domain-Invariant Negative Queue.} In MetaSSL, self-supervised learning is performed within domain-specific tasks, which results in small batch sizes. This poses a challenge when using contrastive learning~\cite{chen2020simple}, a popular self-supervised learning approach, as the objective. In contrastive learning, model performance depends on sufficient negative samples. Thus, it relies on large batch sizes (\textit{e.g.}, 1024) to ensure effective training~\cite{he2020momentum}. However, our MetaSSL involves very small batches (\textit{e.g.}, 128), limiting the available negative samples.

To address this, we propose a \textit{Domain-Invariant Negative Queue} to supplement the small batch sizes across domain-specific tasks. We implement a shared negative queue $\mathcal{N}=\{n_1, n_2, ..., n_k\}$ accessible to each task, where $k$ is defined as a size sufficient for effective contrastive learning (\textit{e.g.}, 1024). When contrastive learning is applied within a task, it draws negatives not only from the current task batch but also from the shared queue $\mathcal{N}$, enriching the pool of negatives for training. Building on Equation~\ref{eq:method_sslloss}, we define the updated contrastive loss function with the Domain-Invariant Negative Queue as follows:
\begin{align}
\notag &\mathcal{L}_\texttt{SSL} (\theta_i; \mathcal{S}_i = \{ x \mid \operatorname{dom}(x) = d_i \}, \mathcal{N}) =\\
\notag &- \sum_{x_j\in \mathcal{S}_i} \log\left( \frac{e^{\text{sim}(f_{\theta_i}(x_j'), f_{\theta_i}(x_j''))}}{e^{\text{sim}(f_{\theta_i}(x_j'), f_{\theta_i}(x_j''))} + Z_j}\right),\\
&\text{where } Z_j= \sum_{k \neq j} e^{\text{sim}(f_{\theta_i}(x_j'), f_{\theta_i}(x_k))} + {\sum_{n \in \mathcal{N}} e^{\text{sim}(f_{\theta_i}(x_j'), f_{\theta_i}(n))}}.
\end{align}

The queue $\mathcal{N}$ must provide effective negative samples across different domain-specific tasks. Hard negatives improve contrastive learning~\cite{he2020momentum}, introducing challenging examples that strengthen model robustness. Thus, we design the elements in $\mathcal{N}$ to function as \textit{domain-invariant hard negatives}. This introduces a challenge, as randomly sampling negatives from the source domain $D^s$ fails to consistently yield hard negatives for different domains, and identifying hard negatives for each domain-specific task is computationally costly.

To overcome this, we define the negative samples in $\mathcal{N}$ as trainable variables~\cite{hu2020adco} and propose an adversarial training approach to optimize them as domain-invariant hard negatives. In this setup, the elements of $\mathcal{N}$ are updated adversarially during each iteration of MetaSSL to maximize domain-specific losses, effectively serving as \textit{general} challenging examples across different domains. The elements in $\mathcal{N}$ are updated when the global weights $\theta$ are updated, following the aggregation of evaluation results from each query set $\mathcal{Q_i}$. The optimization of $\mathcal{N}$ follows a min-max problem:
\begin{equation}
\operatorname{arg} \operatorname*{max}_{\mathcal{N}} \operatorname*{min}_{\theta} \sum_i \mathcal{L}_\texttt{SSL} (\theta_i; \mathcal{Q}_i = \{ x \mid \operatorname{dom}(x) = d_i \}, \mathcal{N}).
\label{eq:negative_queue}
\end{equation}

Our Domain-Invariant Negative Queue enables contrastive learning within each domain-specific task by using an enriched pool of negative samples, even with small batch sizes. Later, in the target-side adaptation step, where target domain data is limited, the trained queue $\mathcal{N}$ is deployed alongside the model to serve as additional negative samples for the adaptation.}

\subsection{Replayed Self-Supervised Learning} \label{sec:replayssl}

\rev{The models pre-trained with MetaSSL are deployed to users, where each user’s domain is considered a target domain $D^t$. Although these models are designed to be domain-adaptive, standard fine-tuning alone fails to adapt them to target domains. Standard fine-tuning relies on few-shot supervised learning, whereas MetaSSL pre-trains models specifically for few-shot self-supervised learning, which does not align with the supervised objective. Consequently, an additional adaptation step is needed to align the pre-trained model with the target domain.

We propose \textit{Replayed Self-Supervised Learning (ReplaySSL)} as the adaptation method for aligning models to target domains. ReplaySSL replays the meta-learned self-supervised task as an adaptation step before supervised fine-tuning using the few-shot data set aside for fine-tuning. This procedure aligns the model more closely with the target domain, creating a \textit{personalized} model.

ReplaySSL adapts the pre-trained weights $\theta$ using the few-shot fine-tuning data $S$ from the target domain $D^t$ with the same self-supervised learning objective used during MetaSSL, $\mathcal{L}_\texttt{SSL}$. This adaptation is completed in a few steps (\textit{e.g.}, 10) with a fixed learning rate $\alpha$, as formulated in the following equation:
\begin{equation}
\theta \gets \theta-\alpha \nabla_{\theta} \mathcal{L}_\texttt{SSL}(\theta;S=\{x \mid \operatorname{dom}(x) = D^t\}) .
\end{equation}
\noindent If $\mathcal{L}_\texttt{SSL}$ is defined as a contrastive learning loss, using the negative queue $\mathcal{N}$ trained from MetaSSL, the following optimization is applied:
\begin{equation}
\theta \gets \theta-\alpha \nabla_{\theta} \mathcal{L}_\texttt{SSL}(\theta;S=\{x \mid \operatorname{dom}(x) = D^t\}, \mathcal{N}).
\end{equation}

Importantly, ReplaySSL is effective only when coupled with MetaSSL. Replaying self-supervised learning without MetaSSL risks damaging the pre-trained model, as training on few-shot data often leads to an overfitted representation and fails to produce a generalizable model. MetaSSL addresses this by making the model adaptive to few-shot self-supervised settings, mitigating overfitting through domain-specific tasks structured within meta-learning. This synergy between MetaSSL and ReplaySSL is the key aspect of \proj{}, and we validate their combined effectiveness through detailed ablation studies in Section~\ref{sec:ablation_metassl_replayssl}.

After ReplaySSL, supervised learning fine-tunes the model on the downstream task. We apply linear evaluation protocol for this fine-tuning: only the classification head is trained, while the encoder remains frozen. This final, fine-tuned model is ready for use in the user application.}

\section{Experiments}
\label{sec:experiments}

\subsection{Datasets}

We evaluate \proj{} on mobile sensing benchmarks across human activity recognition, gesture recognition, and stress level detection tasks. Each dataset has different domains of users or device positions. Our experiments assess performance under domain shift when self-supervised models are fine-tuned on different target domains. The following datasets are used:

\noindent\textbf{ICHAR~\cite{gong2019metasense}} comprises inertial measurement unit (IMU) data for classifying nine types of daily activities, such as walking, running, and stair climbing. Data was collected from ten participants using various mobile devices (seven smartphones and three watches). Each participant is treated as a unique domain.

\noindent\textbf{HHAR~\cite{stisen2015smart}} is designed for classifying six human activities collected from nine users with a combination of four smartwatches and eight smartphones. We define domains based on distinct user-device pairs.

\noindent\textbf{PAMAP2~\cite{reiss2012introducing}} classifies 12 different activity types using data collected from IMUs placed on three body locations: the wrist, chest, and ankle. Domains are divided by device positions.

\noindent\textbf{DSA~\cite{altun2010comparative}} encompasses a wide array of 19 daily and sports activities data, gathered from eight participants wearing five IMUs on the torso, arms, and legs. We define domains based on device positions.

\noindent\textbf{NinaproDB5~\cite{pizzolato2017comparison}} classifies hand gestures using 16-channel Surface Electromyographic (sEMG) signals. We use twelve gestures from Exercise A, focusing on basic finger movements as in prior work~\cite{chen2020hand}. \shepherd{Domains are defined based on individual users (Session~\ref{sec:overall_results}) and session variations (Session~\ref{sec:time_domain}), where sessions represent different recordings that introduce temporal variability.}

\noindent\textbf{WESAD~\cite{schmidt2018introducing}} classifies stress levels across neutral, stress, and amusement states. Physiological and motion data were collected from fifteen participants using wrist- and chest-worn devices. We used only the chest Electrocardiogram (ECG) signals, the most reliable modality for stress detection~\cite{schmidt2018introducing}. Domains are defined based on individual users.

\noindent\shepherd{\textbf{Opportunity~\cite{roggen2010collecting}} captures motion sensor recordings of users performing daily activities. We use accelerometer data from the right wrist to classify four primitive activities: standing, walking, sitting, and lying. To evaluate temporal domain shift, we define domains based on session variations, where each session corresponds to a distinct recording instance.}

\begin{table*}[t]
\centering
\caption{F1-scores of \proj{} and baseline methods for 10-shot fine-tuning across six datasets, with the highest scores in bold and the second-highest underlined.}
\label{tab:eval_baselines}
\fontsize{8pt}{9pt}\selectfont
{\renewcommand{\arraystretch}{1.25}
\begin{tabularx}{0.95\textwidth}{{l}*{8}{C}}
\Xhline{2\arrayrulewidth}
&     & \multicolumn{4}{c}{Domain: User}     & \multicolumn{2}{c}{Domain: Position} &  \\ \cmidrule(lr){3-6} \cmidrule(lr){7-8}
Pre-train & Fine-tune  & ICHAR   & HHAR & NinaproDB5 & WESAD    & PAMAP2  & DSA  & Avg.            \\ \Xhline{2\arrayrulewidth}
\multirow{2}{*}{SimCLR \cite{tang2020exploring}}     & Linear eval. & 0.745\scalebox{0.7}{ $\pm$ 0.024}  & \underline{0.866}\scalebox{0.7}{ $\pm$ 0.008}  &
\underline{0.446}\scalebox{0.7}{ $\pm$ 0.012} &
0.848\scalebox{0.7}{ $\pm$ 0.024} &
0.549\scalebox{0.7}{ $\pm$ 0.016}  & 0.391\scalebox{0.7}{ $\pm$ 0.006} & \underline{0.641}\scalebox{0.7}{ $\pm$ 0.015}          \\
                            & End-to-end   & 0.663\scalebox{0.7}{ $\pm$ 0.028}  & 0.836\scalebox{0.7}{ $\pm$ 0.029}  & 
                            0.405\scalebox{0.7}{ $\pm$ 0.032} &
                            0.869\scalebox{0.7}{ $\pm$ 0.023} & \underline{0.589}\scalebox{0.7}{ $\pm$ 0.046}  & 0.253\scalebox{0.7}{ $\pm$ 0.022} & 0.602\scalebox{0.7}{ $\pm$ 0.030} \\ \hline
\multirow{2}{*}{Set-SimCLR \cite{lee2023self}} & Linear eval. & \underline{0.758}\scalebox{0.7}{ $\pm$ 0.010}  & 0.814\scalebox{0.7}{ $\pm$ 0.004}  & 
0.154\scalebox{0.7}{ $\pm$ 0.012} &
0.813\scalebox{0.7}{ $\pm$ 0.012} &
0.487\scalebox{0.7}{ $\pm$ 0.011}  & 0.283\scalebox{0.7}{ $\pm$ 0.007} & 0.552\scalebox{0.7}{ $\pm$ 0.009}          \\
                            & End-to-end   & 0.747\scalebox{0.7}{ $\pm$ 0.029}  & 0.848\scalebox{0.7}{ $\pm$ 0.016}  &
                            0.244\scalebox{0.7}{ $\pm$ 0.034}  & \underline{0.882}\scalebox{0.7}{ $\pm$ 0.016} &
                            0.573\scalebox{0.7}{ $\pm$ 0.015}  & 0.165\scalebox{0.7}{ $\pm$ 0.012} & 0.577\scalebox{0.7}{ $\pm$ 0.020}          \\ \hline
\multirow{2}{*}{DARLING \cite{zhang2022towards}}    & Linear eval. & 0.749\scalebox{0.7}{ $\pm$ 0.019}  & 0.831\scalebox{0.7}{ $\pm$ 0.003}  & 
0.303\scalebox{0.7}{ $\pm$ 0.013} &
0.789\scalebox{0.7}{ $\pm$ 0.051} &
0.551\scalebox{0.7}{ $\pm$ 0.012}  & \underline{0.399}\scalebox{0.7}{ $\pm$ 0.008} & 0.604\scalebox{0.7}{ $\pm$ 0.018}          \\
                            & End-to-end   & 0.656\scalebox{0.7}{ $\pm$ 0.019}  & 0.844\scalebox{0.7}{ $\pm$ 0.026}  & 
                            0.324\scalebox{0.7}{ $\pm$ 0.028} &
                            0.860\scalebox{0.7}{ $\pm$ 0.030} &
                            0.580\scalebox{0.7}{ $\pm$ 0.042}  & 0.258\scalebox{0.7}{ $\pm$ 0.024} & 0.587\scalebox{0.7}{ $\pm$ 0.028}          \\ \hline
\rowcolor{gray!15} \multicolumn{2}{c}{\proj{} (ours)} & \textbf{0.839\scalebox{0.7}{ $\pm$ 0.023}} & \textbf{0.912\scalebox{0.7}{ $\pm$ 0.009}} & \textbf{0.464\scalebox{0.7}{ $\pm$ 0.021}} &
\textbf{0.883\scalebox{0.7}{ $\pm$ 0.018}} &
\textbf{0.680\scalebox{0.7}{ $\pm$ 0.027}} & \textbf{0.632\scalebox{0.7}{ $\pm$ 0.014}} & \textbf{0.735\scalebox{0.7}{ $\pm$ 0.019}} \\ \Xhline{2\arrayrulewidth}
\end{tabularx}}
\end{table*}

\subsection{Data Preprocessing}

For the human activity recognition datasets, we segmented data using a fixed window size of 256 with an overlap of 128. We standardized each dataset to a range of -1 to 1, following data processing settings from prior work~\cite{gong2019metasense}. We focused exclusively on 3-channel accelerometer data from all sources to reproduce existing baselines~\cite{haresamudram2022assessing} and excluded domains with fewer than 500 samples to ensure sufficient training data. As a result, we used 20 domains for HHAR, representing combinations of five users and four devices. For WESAD, we applied a window size of 10 seconds with a sliding interval of 0.25 seconds, creating approximately 7,000 windows to reduce the computational cost, following a prior study~\cite{bhatti2021attentive}. For NinaproDB5, we used a window size of 60 (0.3 seconds) with a 50\% overlap, following the common practice of using small windows for sEMG data~\cite{josephs2020semg, rahimian2021fs}.

To split data for self-supervised pre-training, fine-tuning, and testing, we followed the approach of previous studies on self-supervised learning for sensing~\cite{haresamudram2022assessing}. \shepherd{Specifically, we allocated 70\% of the data for pre-training---7\% for validation and 63\% for training---and used the remaining 30\% for few-shot fine-tuning. In the few-shot setting, we sampled a few instances per class (\textit{e.g.}, 1, 2, 5, or 10 samples).} The remaining data was split evenly for validation and testing. We ensured no temporal overlap between samples in different splits, preserving data independence.

For our evaluation, we composed pre-training in domains separate from fine-tuning and testing. We prepared fine-tuning and testing sets for each target domain and composed a pre-training dataset by sampling exclusively from other domains. This process was repeated across all target domains to ensure a consistent setup.

\subsection{Baselines}

We benchmark \proj{} against baselines chosen for their efficacy in mitigating domain shift between the unsupervised pre-training and the following fine-tuning. Note that most existing domain generalization~\cite{qian2021latent, qin2022domain, lu2022semantic} and adaptation~\cite{chang2020systematic, gong2019metasense, zhou2020xhar, wang2018stratified} methods assume labeled data for pre-training and thus they do not apply to our scenario. We found two approaches that fit our scenario.

\noindent\textbf{DARLING~\cite{zhang2022towards}} is a domain generalization approach tailored for contrastive learning. DARLING optimizes the loss by using intra-domain negative samples, encouraging discrimination within each domain. This process enables the model to learn domain-invariant features that can be fine-tuned across different domains.

\noindent\textbf{Set-SimCLR~\cite{lee2023self}} is an unsupervised meta-learning method that employs a set encoder to enhance the agreement between augmented sample sets originating from identical sources. Both an instance encoder and the set encoder are trained through contrastive learning. In fine-tuning, the set encoder generates class prototypes from sample sets by class and is used to set the initial weights of the classifier. The classifier, adjusted by the prototypes, facilitates rapid adaptation to novel conditions with the initial weights.




\subsection{Implementation}
\label{sec:implementation_details}

While \proj{} serves as a model- and method-agnostic approach applicable to various self-supervised learning methods, our primary implementation was based on SimCLR~\cite{chen2020simple} to ensure a fair comparison with baselines. This selection aligns with DARLING and Set-SimCLR, which were also presented based on contrastive learning. All baseline models use the same network architecture as \proj{} to maintain consistency.

Our backbone network is implemented with 1D convolutional neural networks (CNNs), followed by a projection head consisting of a fully connected layer. The architecture and hyperparameters are based on SimCLR practices for sensing tasks as described in a previous study~\cite{haresamudram2022assessing}. We optimized hyperparameters through a grid search: for pre-training, we explored learning rates from $\{0.0001, 0.0005, 0.001, 0.005\}$ and weight decays from $\{0, 0.0001\}$. Baseline batch sizes were tested at $\{1024, 2048, 4096\}$, while \proj{} batches were constructed within each domain-specific task, resulting in smaller batch sizes. During fine-tuning, we used a fixed learning rate of $0.005$ for linear evaluation and $0.001$ for end-to-end fine-tuning. Our main evaluation used $\{1, 2, 5, 10\}$-shot samples per class, with additional tests on smaller sample sizes. Using the Adam optimizer, we trained models for 100 epochs in pre-training and 20 epochs in fine-tuning.

For MetaSSL, we optimized the meta-learning rate ($\beta$) and weight decay within the same range as the baselines, with additional tuning for task-specific learning rates ($\alpha$) from $\{0.001, 0.005, 0.01\}$ and inner iteration steps from $\{10, 20, 30\}$. ReplaySSL followed the same parameter setup as MetaSSL. We fixed the number of domain-specific tasks at eight, multi-conditioned tasks at four, and task size at $128$. Since meta-learning requires extended training for convergence, we ran MetaSSL for 5,000 epochs. For the Domain-Invariant Negative Queue, we tested sizes of $\{1024, 2048, 4096\}$, selecting the optimal size to align with the SimCLR baseline batch size. Negative queue elements were optimized using an adversarial objective (Equation~\ref{eq:negative_queue}) with the Adam optimizer and a learning rate of 1. We implemented all methods in PyTorch and conducted training on eight NVIDIA TITAN Xp GPUs.

\subsection{Evaluation Protocols and Metric}

We employed a leave-one-domain-out setting~\cite{qian2021latent}. For each domain in the dataset, we designated it as the target domain for fine-tuning and testing, while all other domains were used for pre-training. This evaluation was conducted across all domains, with each domain rotated as the target, and the results were averaged. We selected a few samples per class (\eg 1, 2, 5, and 10) for fine-tuning and evaluated performance within the same target domain.

We applied two fine-tuning protocols: linear and end-to-end fine-tuning. Linear evaluation served as the primary protocol for \proj{}, treating the pre-trained encoder as a frozen feature extractor and training only a linear classification layer. Since \proj{}’s ReplaySSL refines the encoder parameters, we also conducted end-to-end fine-tuning for baseline methods, updating the entire network without freezing the encoder. This ensured a fair comparison by enabling encoder parameter updates in both cases.

All evaluations were performed using five random seeds, and the results are reported as the mean and standard deviation. To assess performance, we used the macro-averaged F1-score, which is well-suited for handling class imbalances in the data.

\begin{figure*}[t]
    \centering
    \includegraphics[width=0.9\textwidth]{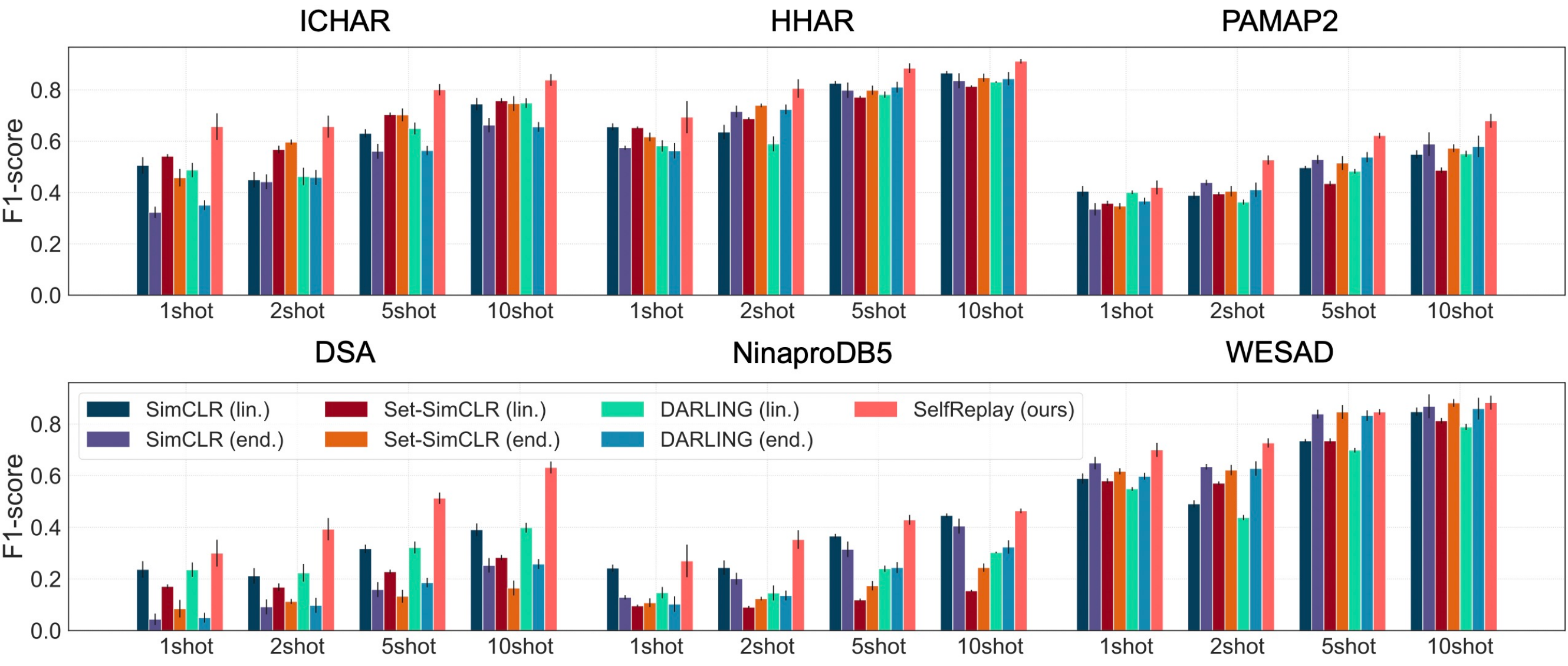}
    \vspace{-8pt}
    \caption{Average F1-scores of \proj{} and the baselines across different shot numbers (1, 2, 5, and 10).}
    \Description{Average F1-scores of \proj{} and the baselines across different shot numbers.}
    \label{fig:baseline_shots}
    \vspace{-5pt}
\end{figure*}

\begin{table*}[t]
\centering
\caption{F1-scores of \proj{} based on different self-supervised learning methods (SimCLR, CPC, and Multi-Task Learning). The highest scores are in bold.}
\vspace{-3pt}
\label{tab:eval_ssl}
\fontsize{8pt}{9pt}\selectfont
{\renewcommand{\arraystretch}{1.25}
\begin{tabularx}{0.8\textwidth}{l *{6}{C}}
\Xhline{2\arrayrulewidth}
&  & \multicolumn{2}{c}{Domain: User} & \multicolumn{2}{c}{Domain: Position} &  \\ \cmidrule(lr){3-4} \cmidrule(lr){5-6}
Pre-train & Fine-tune    & ICHAR            & HHAR             & PAMAP2           & DSA              &   Avg.              \\ \Xhline{2\arrayrulewidth}
\multirow{2}{*}{SimCLR \cite{tang2020exploring}}      & Linear eval. & 0.745\scalebox{0.7}{ $\pm$ 0.024}  & 0.866\scalebox{0.7}{ $\pm$ 0.008}  & 0.549\scalebox{0.7}{ $\pm$ 0.016}  & 0.391\scalebox{0.7}{ $\pm$ 0.006} & 0.638\scalebox{0.7}{ $\pm$ 0.014}\\
                             & End-to-end   & 0.663\scalebox{0.7}{ $\pm$ 0.028}  & 0.836\scalebox{0.7}{ $\pm$ 0.029}  & 0.589\scalebox{0.7}{ $\pm$ 0.046}  & 0.253\scalebox{0.7}{ $\pm$ 0.022}  & 0.585\scalebox{0.7}{ $\pm$ 0.031}\\  
                            \rowcolor{gray!15} \multicolumn{2}{c}{\proj$_{\texttt{SimCLR}}$ (ours)} & \textbf{0.839\scalebox{0.7}{ $\pm$ 0.023}} & \textbf{0.912\scalebox{0.7}{ $\pm$ 0.009}} & \textbf{0.680\scalebox{0.7}{ $\pm$ 0.027}} & \textbf{0.632\scalebox{0.7}{ $\pm$ 0.014}} & \textbf{0.735\scalebox{0.7}{ $\pm$ 0.019}} \\ \hline
\multirow{2}{*}{CPC \cite{haresamudram2023investigating}}         & Linear eval. & 0.765\scalebox{0.7}{ $\pm$ 0.016}  & 0.846\scalebox{0.7}{ $\pm$ 0.005}  & 0.379\scalebox{0.7}{ $\pm$ 0.017}  & 0.371\scalebox{0.7}{ $\pm$ 0.005} & 0.590\scalebox{0.7}{ $\pm$ 0.011} \\
                             & End-to-end   & 0.816\scalebox{0.7}{ $\pm$ 0.013}  & 0.849\scalebox{0.7}{ $\pm$ 0.021}  & 0.484\scalebox{0.7}{ $\pm$ 0.026}  & 0.352\scalebox{0.7}{ $\pm$ 0.017} & 0.625\scalebox{0.7}{ $\pm$ 0.019}\\ 
                             \rowcolor{gray!15} \multicolumn{2}{c}{\proj$_{\texttt{CPC}}$ (ours)} & \textbf{0.826\scalebox{0.7}{ $\pm$ 0.008}} & \textbf{0.871\scalebox{0.7}{ $\pm$ 0.005}} & \textbf{0.527\scalebox{0.7}{ $\pm$ 0.017}} & \textbf{0.419\scalebox{0.7}{ $\pm$ 0.008}} & \textbf{0.661\scalebox{0.7}{ $\pm$ 0.009}} \\ \hline
\multirow{2}{*}{Multi-Task \cite{saeed2019multi}}  & Linear eval. & 0.716\scalebox{0.7}{ $\pm$ 0.010}  & 0.877\scalebox{0.7}{ $\pm$ 0.003}  & 0.630\scalebox{0.7}{ $\pm$ 0.003}  & 0.456\scalebox{0.7}{ $\pm$ 0.004} & 0.670\scalebox{0.7}{ $\pm$ 0.005} \\
                             & End-to-end   & 0.718\scalebox{0.7}{ $\pm$ 0.019} & 0.865\scalebox{0.7}{ $\pm$ 0.030} & 0.636\scalebox{0.7}{ $\pm$ 0.015} & 0.378\scalebox{0.7}{ $\pm$ 0.021}  & 0.649\scalebox{0.7}{ $\pm$ 0.021} \\ 
                             \rowcolor{gray!15} \multicolumn{2}{c}{\proj$_{\texttt{MultiTask}}$ (ours)} & \textbf{0.794\scalebox{0.7}{ $\pm$ 0.015}} & \textbf{0.891\scalebox{0.7}{ $\pm$ 0.005}} & \textbf{0.659\scalebox{0.7}{ $\pm$ 0.016}} & \textbf{0.578\scalebox{0.7}{ $\pm$ 0.011}} & \textbf{0.731\scalebox{0.7}{ $\pm$ 0.012}} \\ \Xhline{2\arrayrulewidth}
\end{tabularx}}
\end{table*}

\begin{figure*}[t]
    \centering
    \includegraphics[width=0.85\textwidth]{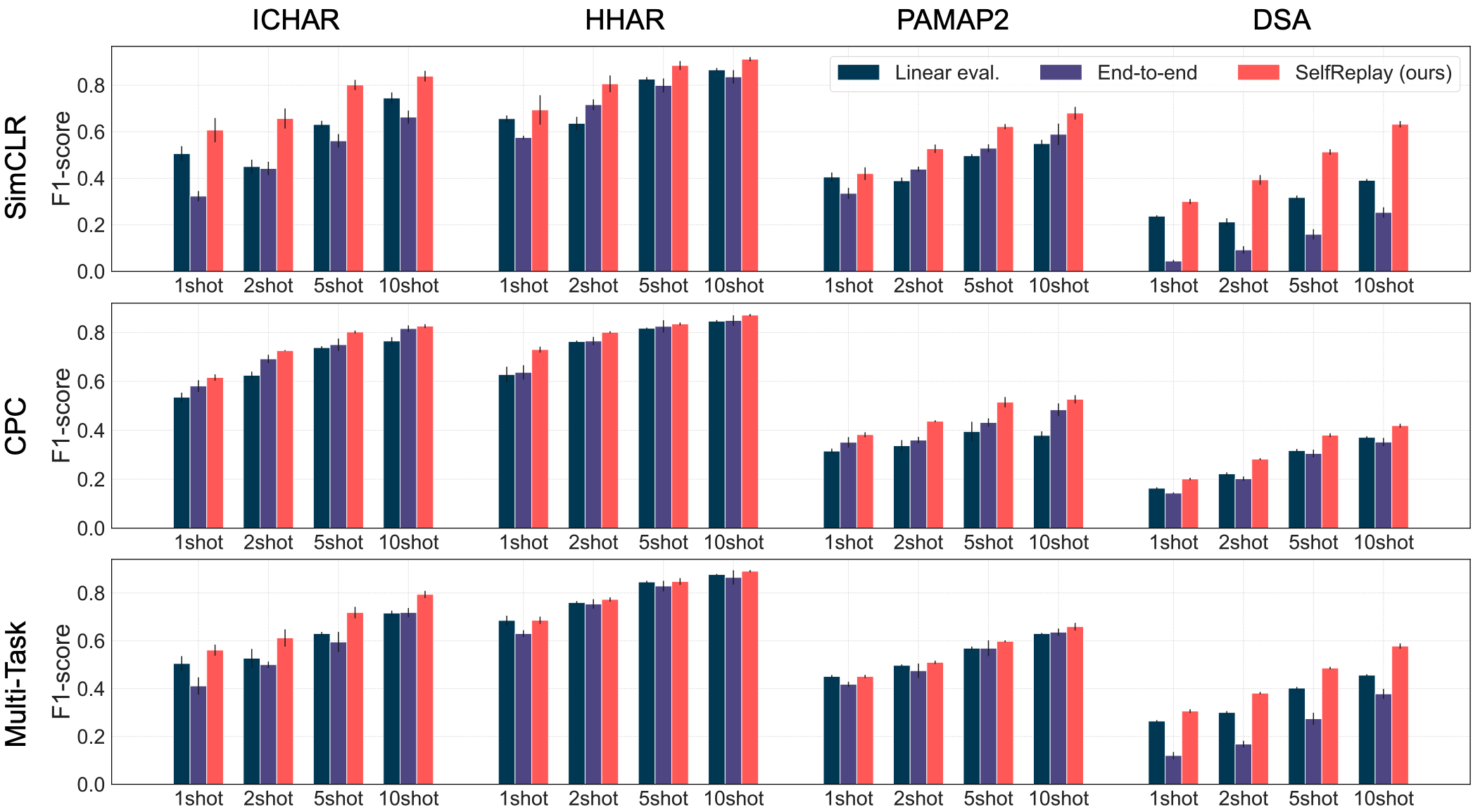}
    \vspace{-5pt}
    \caption{Average F1-scores of \proj{} based on different self-supervised learning methods (SimCLR, CPC, and multi-task learning) across different shot numbers (1, 2, 5, and 10).}
    \Description{Average F1-scores of \proj{} based on different self-supervised learning methods.}
    \label{fig:ssl_shots}
    \vspace{-8pt}
\end{figure*}

\subsection{Results}

\subsubsection{Main Evaluation}
\label{sec:overall_results}

Table~\ref{tab:eval_baselines} shows the performance of \proj{} compared with domain generalization and domain adaptation baselines, using 10-shot fine-tuning. Bolded values indicate the highest scores in each column. \proj{} consistently achieves the highest F1 scores across all datasets. The results indicate that baseline models struggle to capture domain-specific features as they rely on weights pre-trained in the source domain. While end-to-end fine-tuning occasionally improves performance, its impact varies widely depending on the dataset, likely due to few-shot fine-tuning’s sensitivity. Baselines do not fully benefit from end-to-end fine-tuning without an adaptive pre-training design. In contrast, \proj{} achieves an average F1-score improvement of 9.4\%p, establishing it as a robust domain adaptation method.

\subsubsection{Performance across Different Shots}

\shepherd{We evaluated the performance of \proj{} compared with the baselines by fine-tuning with a smaller number of samples (\textit{e.g.}, 1, 2, and 5 shots per class).} Figure~\ref{fig:baseline_shots} shows the results. For all datasets and shot settings, \proj{} consistently achieves the best performance, with improvements of 5.9\%, 14.3\%, 12.1\%, and 9.4\% over the second-best method, respectively. These results demonstrate the robustness of our approach across varying few-shot scenarios, indicating that \proj{} remains effective in data-scarce mobile sensing applications.

\subsubsection{Integration with Self-Supervised Learning Methods}
\label{sec:ssl_results}

We designed \proj{} to be agnostic to specific self-supervised learning methods. For evaluation, we implemented \proj{} with two additional self-supervised objectives: contrastive predictive coding (CPC)~\cite{oord2018representation} (\proj{}$_\texttt{CPC}$) and multi-task learning~\cite{saeed2019multi} (\proj{}$_\texttt{MultiTask}$). Each version replaces the self-supervised loss function $\mathcal{L}_\texttt{SSL}$ (from Equation~\ref{eq:method_sslloss}) with the objective corresponding to its respective method, CPC or multi-task learning. CPC defines an objective for predicting the embedding of a future window based on past embeddings, contrasting the true future window against other candidates. Multi-task learning assigns the model several tasks focused on identifying data transformations and promoting shared feature learning across tasks.

In implementing \proj{}$_\texttt{CPC}$ and \proj{}$_\texttt{MultiTask}$ along with their baselines, we followed the same architecture and parameter tuning settings from prior assessments~\cite{haresamudram2022assessing}. For parameter tuning, we used the same search settings as outlined in Section~\ref{sec:implementation_details}. The only adjustment was in batch size, which was reduced for CPC and multi-task learning, searching across $\{64, 128, 256\}$ for optimal results. Since CPC and multi-task learning were primarily designed for human activity recognition~\cite{haresamudram2022assessing}, we focused our evaluations on four relevant datasets.

Table~\ref{tab:eval_ssl} presents the results across different self-supervised learning methods. \proj{} consistently improved performance with average gains of 4.4\%p and 6.1\%p for CPC and multi-task learning, respectively. Performance gains varied depending on the self-supervised method used on the same dataset, which we analyze further in Section~\ref{sec:domain_shift_effect}. Additionally, Figure~\ref{fig:ssl_shots} illustrates \proj{}’s effectiveness across $k$-shot settings, underscoring its robustness. These results demonstrate that \proj{} can be applied flexibly across different applications, adapting to the most effective self-supervised learning method for each setting.

\begin{table}[t]
\centering
\caption{\shepherd{F1-scores of \proj{} and baseline methods for 10-shot fine-tuning on two datasets, where domains are defined by sessions to evaluate temporal domain shift. The highest scores are in bold and the second-highest are underlined.}}
\vspace{-4pt}
\label{tab:time_domain}
\fontsize{8pt}{9pt}\selectfont
{\renewcommand{\arraystretch}{1.25}
\begin{tabularx}{0.95\columnwidth}{{l}*{3}{C}}
\Xhline{2\arrayrulewidth}
&     & \multicolumn{2}{c}{Domain: Session} \\
\cmidrule(lr){3-4}
Pre-train & Fine-tune & Opportunity & NinaproDB5 \\ \Xhline{2\arrayrulewidth}
\multirow{2}{*}{SimCLR \cite{tang2020exploring}}     & Linear eval. & 0.427\scalebox{0.7}{ $\pm$ 0.007}  & 0.676\scalebox{0.7}{ $\pm$ 0.006} \\
                            & End-to-end   & \underline{0.458}\scalebox{0.7}{ $\pm$ 0.001}  & \underline{0.693}\scalebox{0.7}{ $\pm$ 0.004} \\ \hline
\multirow{2}{*}{Set-SimCLR \cite{lee2023self}} & Linear eval. & 0.438\scalebox{0.7}{ $\pm$ 0.009}  & 0.206\scalebox{0.7}{ $\pm$ 0.003} \\
                            & End-to-end   & 0.445\scalebox{0.7}{ $\pm$ 0.009}  & 0.287\scalebox{0.7}{ $\pm$ 0.007} \\ \hline
\multirow{2}{*}{DARLING \cite{zhang2022towards}}    & Linear eval. & 0.424\scalebox{0.7}{ $\pm$ 0.004}  & 0.657\scalebox{0.7}{ $\pm$ 0.005} \\
                            & End-to-end   & 0.456\scalebox{0.7}{ $\pm$ 0.010}  & 0.686\scalebox{0.7}{ $\pm$ 0.003} \\ \hline
\rowcolor{gray!15} \multicolumn{2}{c}{\proj{} (ours)} & \textbf{0.476\scalebox{0.7}{ $\pm$ 0.008}} & \textbf{0.724\scalebox{0.7}{ $\pm$ 0.003}} \\ \Xhline{2\arrayrulewidth}
\end{tabularx}}
\vspace{-10pt}
\end{table}

\subsubsection{Robustness to Temporal Domain Shift}
\label{sec:time_domain}

\shepherd{We evaluate an extended domain scope by testing \proj{} on temporal shifts within the same user and device. In this setting, \textit{sessions} serve as temporal domains, capturing changes in user behavior and environmental conditions over time~\cite{shi2022deep}.

We used two datasets, Opportunity and NinaproDB5, providing session separation with distinct measurement times. Since we focus on sessions within a single user, the available pre-training data was limited. To address this, we reduced the window generation step size to 12.5\% and expanded the batch size search space to $\{64, 128, 256, 1024, 2048, 4096\}$ to find an optimal configuration. Experiments were conducted on a randomly selected user.

Table~\ref{tab:time_domain} shows \proj{} outperformed all baselines on both datasets. In NinaproDB5, using sessions as domains yielded higher performance than using users as domains (Section~\ref{sec:overall_results}). Still, temporal shifts persisted, and \proj{} demonstrated the highest robustness. Our findings indicate that \proj{} can be effectively deployed in continuous sensing systems, dynamically adapting to temporal changes through fine-tuning with recent data.}

\begin{figure}
    \centering
    \includegraphics[width=0.9\columnwidth]{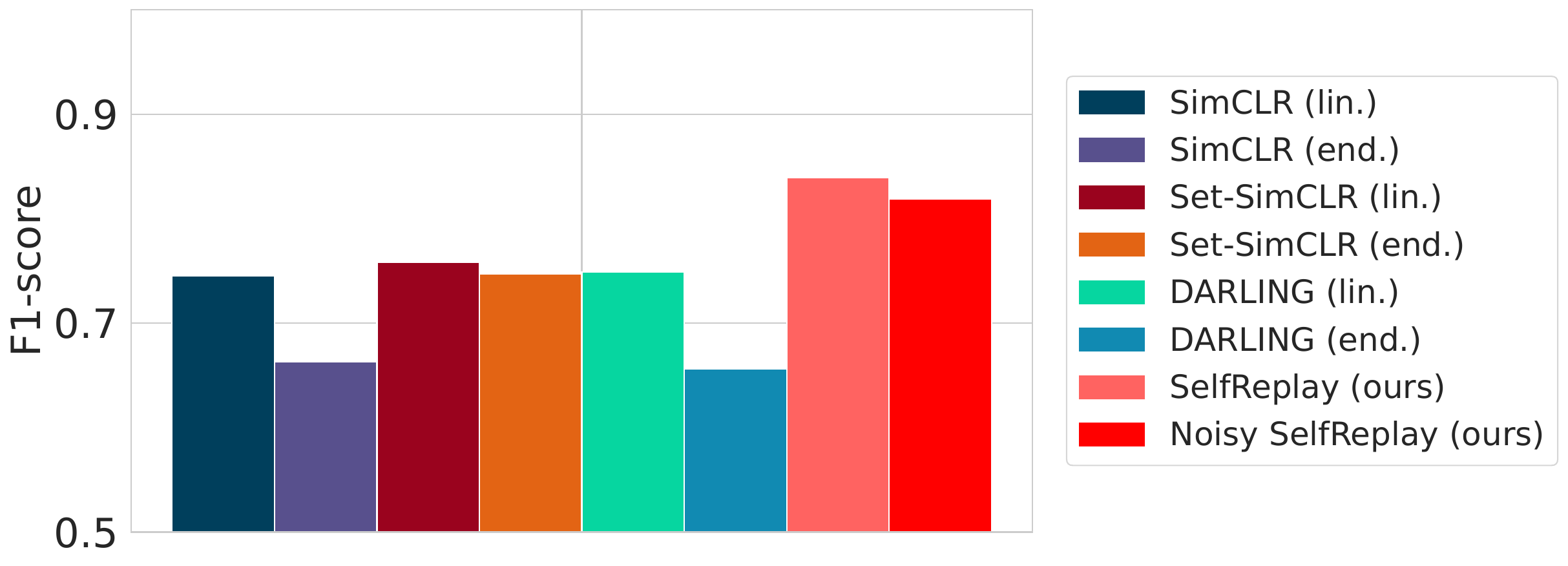}
    \vspace{-5pt}
    \caption{\shepherd{Average F1-scores of \proj{} and baselines with and without noisy domain labels (50\%) during pre-training.}}
    \label{fig:noise_sensitivity}
    \Description{A comparison between the standard pre-training and fine-tuning (top) and \proj{} (bottom).}
    \vspace{-10pt}
\end{figure}

\subsubsection{Robustness to Noisy Domain Labels}

\shepherd{We assume that MetaSSL has access to domain labels during pre-training. To assess the impact of domain labels, we introduced noise into 50\% of the domain labels by assigning incorrect values and evaluated MetaSSL on the ICHAR dataset with SimCLR as the base self-supervised method.

\begin{table*}[t]
\centering
\caption{F1-score comparison of \proj{} against 1) the baseline self-supervised learning method, 2) \proj{} without meta-learning, and 3) \proj{} without replay. 10-shot fine-tuning is performed.}
\vspace{-3pt}
\label{tab:eval_ablation_main}
\fontsize{8pt}{9pt}\selectfont
{\renewcommand{\arraystretch}{1.25}
\begin{tabularx}{0.95\textwidth}{*{2}{C}|*{7}{C}}
\Xhline{2\arrayrulewidth} 
\multicolumn{2}{c|}{Components} & \multicolumn{4}{c}{Domain: User}     & \multicolumn{2}{c}{Domain: Position} &  \\ \cmidrule(lr){1-2} \cmidrule(lr){3-6} \cmidrule(lr){7-8}
MetaSSL & ReplaySSL & ICHAR   & HHAR & NinaproDB5 & WESAD    & PAMAP2  & DSA  & Avg. \\ \Xhline{2\arrayrulewidth}
\ding{55} & \ding{55}
& 0.745\scalebox{0.7}{ $\pm$ 0.024}  
& 0.866\scalebox{0.7}{ $\pm$ 0.008}  
& 0.446\scalebox{0.7}{ $\pm$ 0.012} 
& 0.848\scalebox{0.7}{ $\pm$ 0.024} 
& 0.549\scalebox{0.7}{ $\pm$ 0.016}  
& 0.391\scalebox{0.7}{ $\pm$ 0.006} 
& 0.641\scalebox{0.7}{ $\pm$ 0.014}          
\\ 
\ding{55} & \Checkmark
& 0.731\scalebox{0.7}{ $\pm$ 0.029}  
& 0.638\scalebox{0.7}{ $\pm$ 0.044}  
& 0.436\scalebox{0.7}{ $\pm$ 0.018} 
& 0.853\scalebox{0.7}{ $\pm$ 0.021} 
& 0.464\scalebox{0.7}{ $\pm$ 0.063}  
& 0.294\scalebox{0.7}{ $\pm$ 0.011} 
& 0.569\scalebox{0.7}{ $\pm$ 0.037}          
\\ 
\Checkmark & \ding{55}
& 0.792\scalebox{0.7}{ $\pm$ 0.047}  
& 0.880\scalebox{0.7}{ $\pm$ 0.022}  
& \textbf{0.464\scalebox{0.7}{ $\pm$ 0.021}}
& 0.882\scalebox{0.7}{ $\pm$ 0.018} 
& 0.677\scalebox{0.7}{ $\pm$ 0.021}  
& 0.620\scalebox{0.7}{ $\pm$ 0.012} 
& 0.719\scalebox{0.7}{ $\pm$ 0.026}          
\\
\Checkmark & \Checkmark
& \textbf{0.839\scalebox{0.7}{ $\pm$ 0.023}} 
& \textbf{0.912\scalebox{0.7}{ $\pm$ 0.009}} 
& \textbf{0.464\scalebox{0.7}{ $\pm$ 0.021}} 
& \textbf{0.883\scalebox{0.7}{ $\pm$ 0.018}} 
& \textbf{0.680\scalebox{0.7}{ $\pm$ 0.027}} 
& \textbf{0.632\scalebox{0.7}{ $\pm$ 0.014}} 
& \textbf{0.735\scalebox{0.7}{ $\pm$ 0.019}} 
\\ 
\Xhline{2\arrayrulewidth}
\end{tabularx}}
\end{table*}

\begin{table*}[t]
\centering
\caption{F1-score comparison of \proj{} (w/ Negative Queue) against the baseline self-supervised learning method and \proj{} without the Domain-Invariant Negative Queue (w/o Negative Queue). Results are based on 10-shot fine-tuning.}
\vspace{-3pt}
\label{tab:ablation_neg_queue}
\fontsize{8pt}{9pt}\selectfont
{\renewcommand{\arraystretch}{1.25}
\begin{tabularx}{0.95\textwidth}{{c}*{7}{C}}
\Xhline{2\arrayrulewidth} 
 & \multicolumn{4}{c}{Domain: User}     & \multicolumn{2}{c}{Domain: Position} &  \\ \cmidrule(lr){2-5} \cmidrule(lr){6-7}
 & ICHAR   & HHAR & NinaproDB5 & WESAD    & PAMAP2  & DSA  & Avg.            \\ \Xhline{2\arrayrulewidth}
Baseline 
& 0.745\scalebox{0.7}{ $\pm$ 0.024}  
& 0.866\scalebox{0.7}{ $\pm$ 0.008}  
& 0.446\scalebox{0.7}{ $\pm$ 0.012} 
& 0.848\scalebox{0.7}{ $\pm$ 0.024} 
& 0.549\scalebox{0.7}{ $\pm$ 0.016}  
& 0.391\scalebox{0.7}{ $\pm$ 0.006} 
& 0.641\scalebox{0.7}{ $\pm$ 0.014}          
\\ \hline
w/o Negative Queue 
& 0.836\scalebox{0.7}{ $\pm$ 0.011}  
& 0.903\scalebox{0.7}{ $\pm$ 0.004}  
& 0.449\scalebox{0.7}{ $\pm$ 0.018}
& 0.879\scalebox{0.7}{ $\pm$ 0.020} 
& 0.639\scalebox{0.7}{ $\pm$ 0.030}  
& 0.526\scalebox{0.7}{ $\pm$ 0.019} 
& 0.705\scalebox{0.7}{ $\pm$ 0.017}          
\\
w/ Negative Queue 
& \textbf{0.839\scalebox{0.7}{ $\pm$ 0.023}} 
& \textbf{0.912\scalebox{0.7}{ $\pm$ 0.009}} 
& \textbf{0.464\scalebox{0.7}{ $\pm$ 0.021}} 
& \textbf{0.883\scalebox{0.7}{ $\pm$ 0.018}} 
& \textbf{0.680\scalebox{0.7}{ $\pm$ 0.027}} 
& \textbf{0.632\scalebox{0.7}{ $\pm$ 0.014}} 
& \textbf{0.735\scalebox{0.7}{ $\pm$ 0.019}} 
\\ \Xhline{2\arrayrulewidth}
\end{tabularx}}
\end{table*}

Figure~\ref{fig:noise_sensitivity} shows that with noisy domain labels, \proj{}'s F1 score decreased modestly from 0.839 to 0.819, indicating that label quality does influence performance. Nonetheless, \proj{} still outperformed baseline methods, whose highest score was 0.758. These findings suggest that while accurate domain labels are preferable, the meta-learning framework based on small meta-tasks keeps \proj{} robust even under noisy conditions.}

\begin{figure}
    \centering
    \includegraphics[width=0.9\columnwidth]{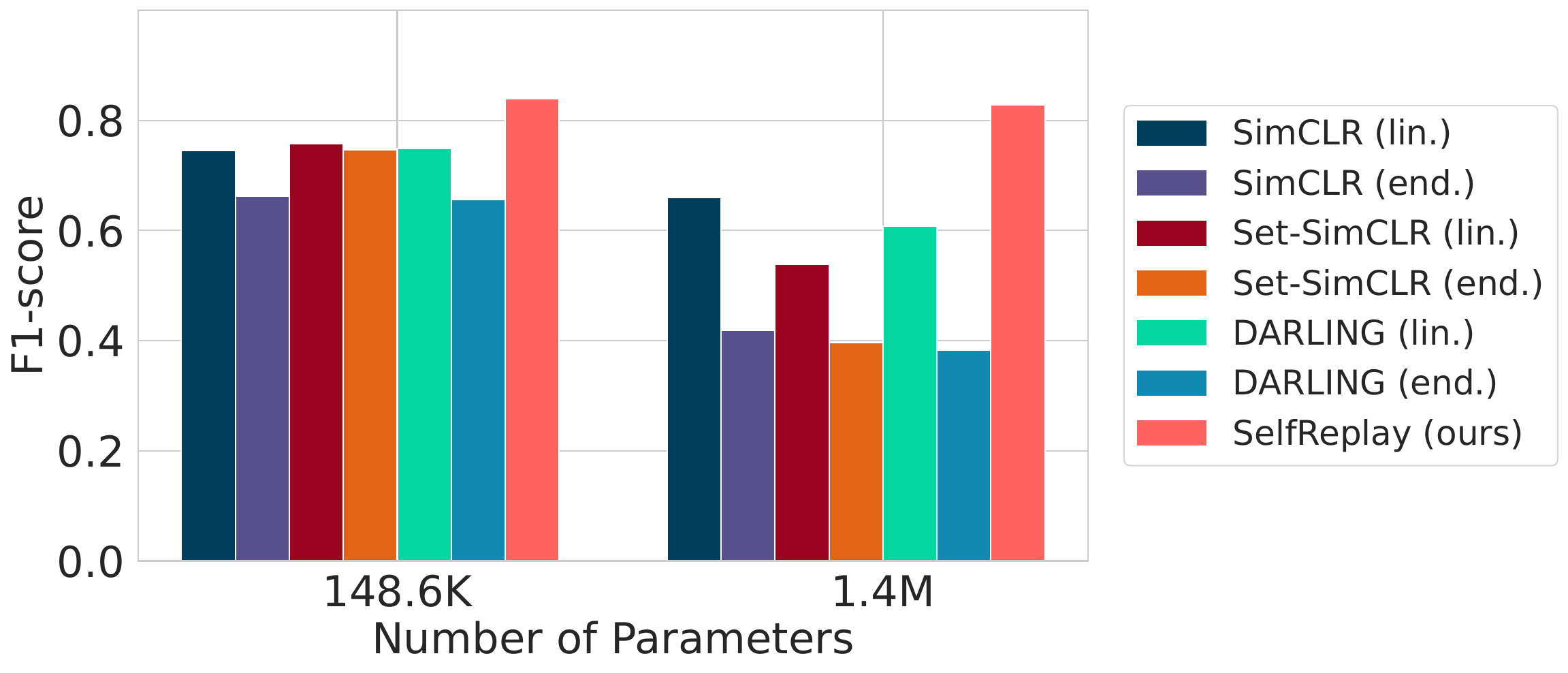}
    \vspace{-5pt}
    \caption{\shepherd{Average F1-scores of \proj{} and baselines for a small (148.6K) model versus a large (1.4M) model.}}
    \label{fig:model_size}
    \Description{A comparison between the standard pre-training and fine-tuning (top) and \proj{} (bottom).}
    \vspace{-5pt}
\end{figure}

\subsubsection{Robustness to Model Size}

\shepherd{
To assess how larger models behave under domain shift, we evaluated the effect of model size in our setting by increasing the number of layers in our model, which expanded the parameter count from 148.6K to 1.4M. Figure~\ref{fig:model_size} illustrates how larger models affect fine-tuning performance in a new domain compared with smaller models on the ICHAR dataset.

Our findings reveal a notable trend: larger models exhibited increased overfitting to the source domain for all baselines, leading to worse fine-tuning performance on the target domain. In contrast, \proj{} maintained the performance, demonstrating robustness to domain shift even in a large model. 

These results suggest that \proj{} effectively mitigates domain shift regardless of model size. In particular, as foundation models for sensing applications are expected to be large, our findings indicate that \proj{} can enhance their generalizability across domains.}

\subsubsection{Ablation Study: MetaSSL and ReplaySSL}
\label{sec:ablation_metassl_replayssl}

We conducted an ablation study to examine the contributions of the key components of \proj{}: MetaSSL and ReplaySSL. We assessed their individual and combined impact by comparing \proj{}'s performance with and without these components.

Table~\ref{tab:eval_ablation_main} shows the results. A key observation is that using ReplaySSL alone, without MetaSSL, leads to a significant performance drop, even falling below the baseline. This suggests that applying self-supervised learning on a limited target domain dataset can lead to overfitting, distorting the pre-trained model weights. This finding underscores MetaSSL’s role as the essential component that enables effective adaptation through ReplaySSL.

With MetaSSL, \proj{} outperforms the baseline, even without ReplaySSL. This result indicates that self-supervised meta-learning promotes meaningful feature learning, which we attribute to meta-learning’s capacity to capture transferable features that support effective adaptation. By effectively “learning to learn,” MetaSSL enhances the model’s ability to leverage generalizable features across domains, which provides a foundation for rapid adaptation with limited data.

Combining MetaSSL and ReplaySSL ultimately achieves the best performance, showing that the adaptation step further enhances the model. This highlights the synergistic value of using MetaSSL and ReplaySSL together.

\begin{figure*}[t]
    \centering
    \includegraphics[width=0.85\textwidth]{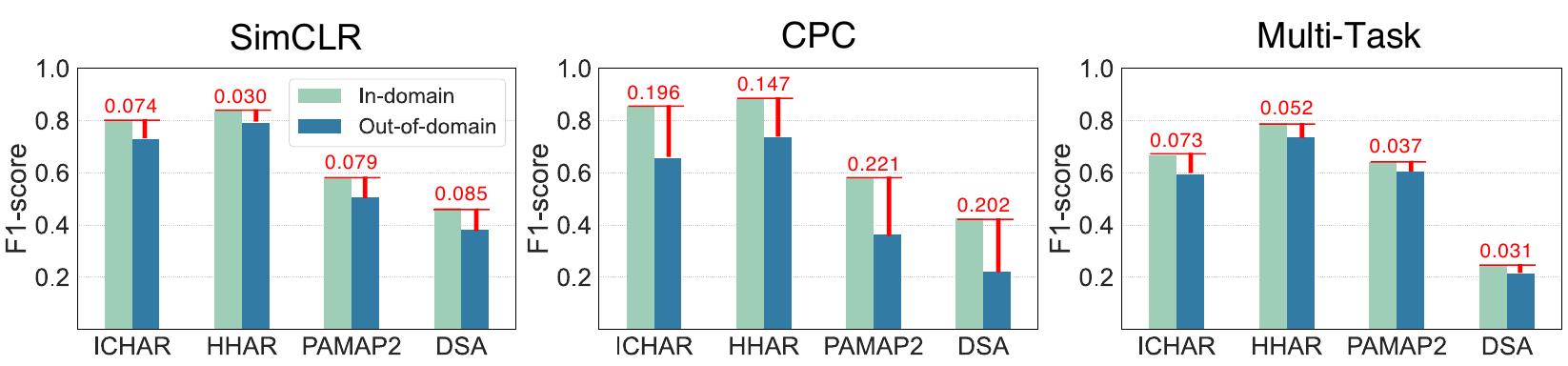}
    \vspace{-7pt}
    \caption{Fine-tuning performance comparison between models pre-trained in-domain and out-of-domain settings. 10-shot fine-tuning is performed for all settings. Performance drops between the settings are shown in red.}
    \Description{Fine-tuning performance comparison between models pre-trained in-domain and out-of-domain settings.}
    \label{fig:domain_effect}
    \vspace{-5pt}
\end{figure*}

\subsubsection{Ablation Study: Domain-Invariant Negative Queue}
\label{sec:ablation_negative_queue}

We conducted an ablation study to assess the impact of Domain-Invariant Negative Queue in \proj{}$_\texttt{SimCLR}$. Without the domain-invariant negative queue, negative samples are drawn exclusively from each domain-specific task, resulting in a limited set of negatives. By contrast, using the negative queue allows access to over 1024 diverse negative samples.

Table~\ref{tab:ablation_neg_queue} shows the results. Although performance improves without the negative queue relative to the baseline (SimCLR), the Domain-Invariant Negative Queue further enhances performance, with an average improvement of 3\%p. This result indicates the effectiveness of the negative queue, as it mitigates the limitations of small batch sizes within \proj{}.

\subsubsection{Domain Effect of Self-Supervised Learning Methods} \label{sec:domain_shift_effect}

Our evaluation of \proj{} across different self-supervised learning methods (Section~\ref{sec:ssl_results}) revealed varying impacts depending on the method. To investigate this difference, we examined how domain shifts affect each method. We pre-trained models using SimCLR, CPC, and multi-task learning in a leave-one-domain-out setup and fine-tuned them in a novel domain (\textit{out-of-domain}) setting. For comparison, we created an \textit{in-domain} baseline, where pre-training and fine-tuning occur within the same domain. We measure each method’s sensitivity to domain shifts by comparing performance drops from in-domain with out-of-domain. We equalized data sizes across both settings to ensure fair comparisons.

Figure~\ref{fig:domain_effect} illustrates the results. Each self-supervised method experiences a drop in performance when fine-tuned and tested on out-of-domain data, highlighting the challenge of domain shifts. However, the level of decline differs by method: CPC shows the largest drop, with an average F1-score reduction of 19.15\%p, while SimCLR and multi-task learning have moderate decreases of 6.7\%p and 4.95\%p, respectively.

These results indicate that the self-supervised learning method influences a model’s sensitivity to domain shifts. CPC, which trains the model to predict future segments, often learns patterns specific to the source domain. For example, a model pre-trained from younger and active users might focus on a pattern of increasing activity over time. When transferred to a domain with older users, whose activity decreases over time, the CPC model struggles because its predictive patterns do not align with the new domain's temporal dynamics. In contrast, multi-task learning's objective of identifying data augmentations is less domain-specific. This knowledge transfers more effectively between user groups, as recognizing augmented characteristics (\eg{} rotation) is independent of specific user trends.

In summary, domain shifts generally lead to performance drops during fine-tuning, with the degree of decline varying by a self-supervised method. Our findings underscore the importance of selecting suitable self-supervised methods for effective deployment in heterogeneous mobile environments. We also stress that self-supervised learning methods consider the effects of domain shift across different settings.

\subsubsection{MetaSSL Overhead}

\shepherd{Unlike standard self-supervised learning methods that process the entire dataset in each epoch, MetaSSL trains on small meta-tasks per epoch (\textit{e.g.}, 12 tasks with a size of 128). As a result, the model observes only a limited portion of the dataset at a time. Therefore, we increased the number of epochs to 5000, which is higher than the standard SSL setting of 100 epochs, to ensure MetaSSL observes enough data and converges. In this setting, we measured the execution time and VRAM usage for MetaSSL and standard SSL methods using a single NVIDIA TITAN XP GPU. Experiments were conducted using the ICHAR dataset. 

Although MetaSSL trains smaller data per iteration, its overall training time was longer due to the additional overhead of meta-task construction, aggregation, and serial task-specific gradient updates following the MAML implementation. As a result, the total runtime increased from 3.7 minutes to 3.3 hours. VRAM usage increased slightly from 1.5GB to 1.6GB, as MetaSSL maintains gradients across multiple meta-tasks rather than just within mini-batches. The overhead was more pronounced for CPC, where the runtime increased from 23 minutes to 23 hours, and VRAM usage grew from 0.94GB to 7.7GB. This was due to CPC's reliance on large embedding variables, which MetaSSL maintains across meta-tasks. For multi-task learning, the total runtime increased from 10.9 minutes to 4.3 hours, and VRAM usage grew from 0.4GB to 1GB.

The source of overhead is our current MAML-based implementation, which is straightforward but processes meta-tasks sequentially. It is important to note that MetaSSL is conducted on servers where pre-training benefits from abundant resources. Meanwhile, to further optimize training costs, more efficient meta-learning methods such as Reptile~\cite{nichol2018reptile} or parallelized meta-task updates could improve efficiency. Additionally, refining task generation strategies may further reduce computational overhead.}

\begin{figure}[t]
    \centering
    \begin{minipage}[t]{0.4\columnwidth}
        \centering
        \includegraphics[width=\columnwidth]{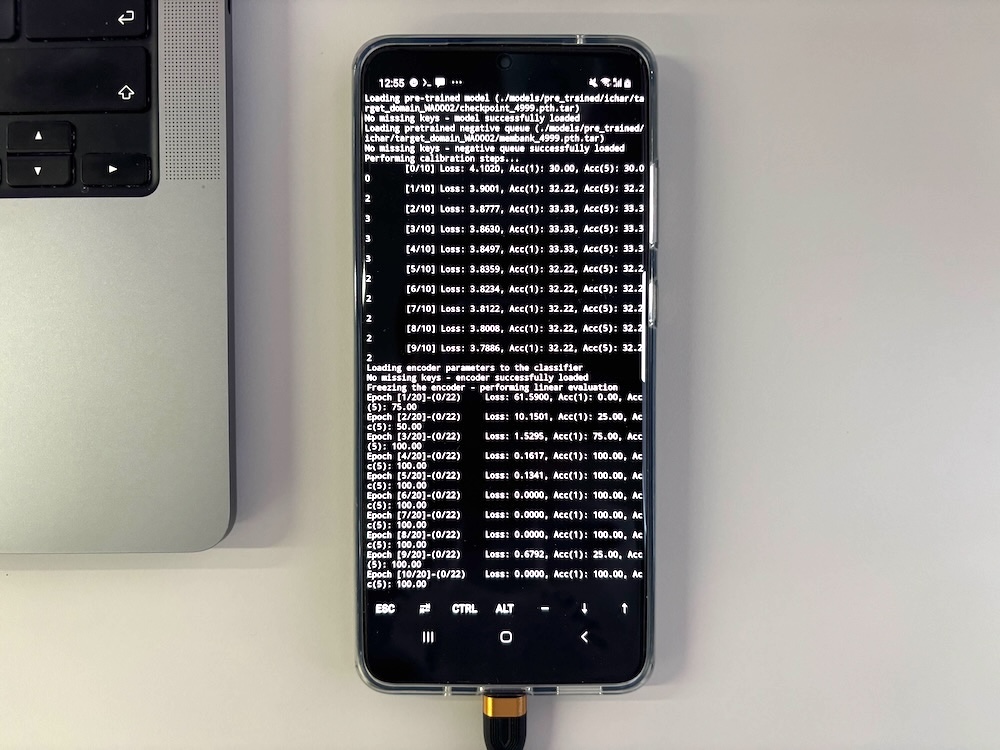}
    \end{minipage}
    \hspace{3pt}
    \begin{minipage}[t]{0.4\columnwidth}
        \centering
        \includegraphics[width=\columnwidth]{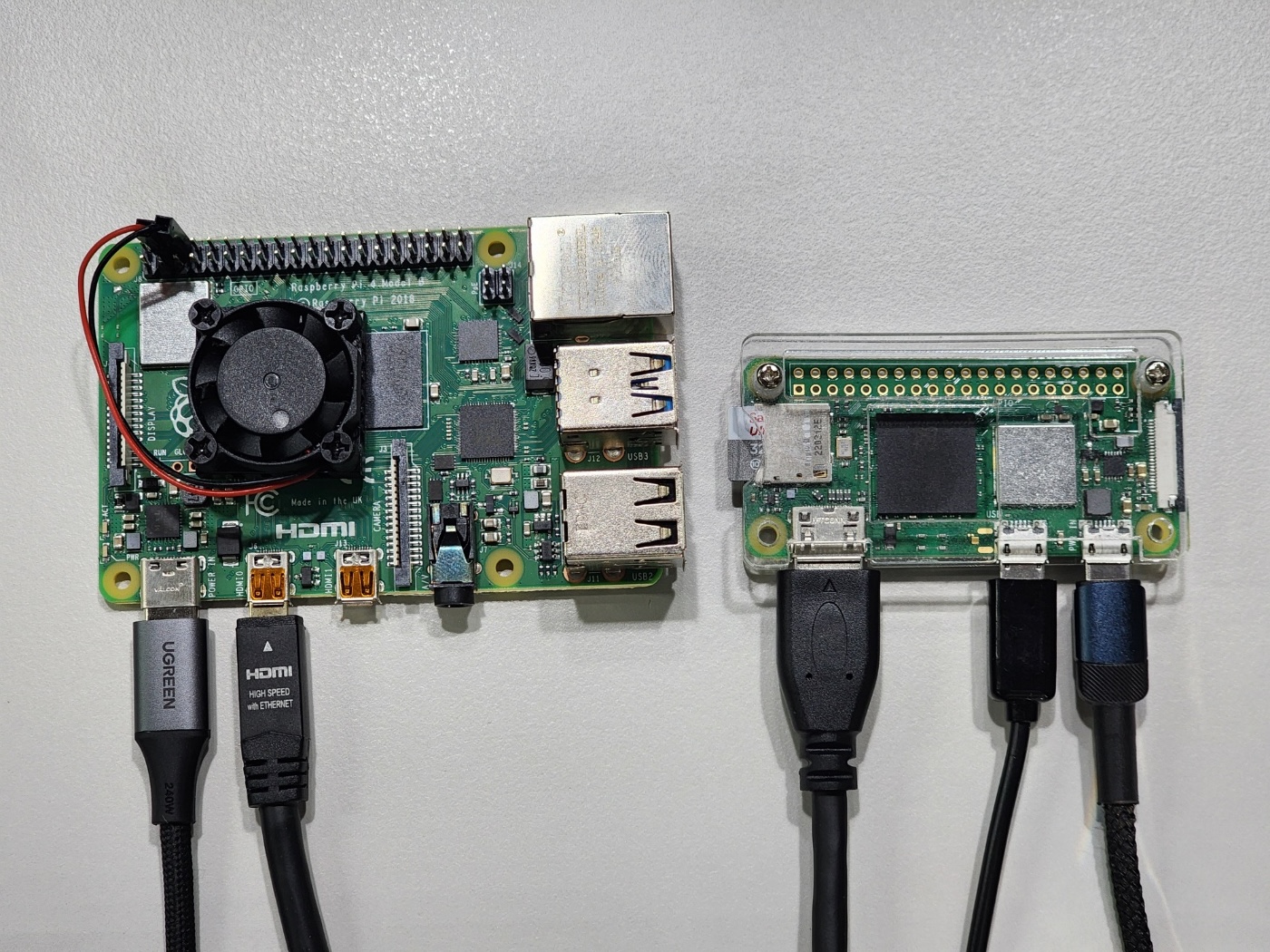}
    \end{minipage}
    \vspace{-3pt}
    \caption{\shepherd{Testbed setups for on-device execution of \proj{} on three devices: Samsung Galaxy S20 Ultra (12GB RAM, left), Raspberry Pi 4 (4GB RAM, middle), and Raspberry Pi Zero 2 W (512MB RAM, right).}}
    \Description{Testbed setups for on-device execution of \proj{} on three devices: Samsung Galaxy S20 Ultra (12GB RAM, left), Raspberry Pi 4 (4GB RAM, middle), and Raspberry Pi Zero 2 W (512MB RAM, right).}
    \label{fig:ondevice}
    \vspace{-10pt}
\end{figure}

\begin{table}[t]
\small{
\centering
\caption{\shepherd{Computational overhead of ReplaySSL followed by fine-tuning across different self-supervised learning methods on three edge devices: Samsung Galaxy S20 Ultra, Raspberry Pi 4, and Raspberry Pi Zero 2 W.}}
\vspace{-3pt}
\label{tab:overhead}
{\renewcommand{\arraystretch}{1.1}
\begin{tabularx}{1\columnwidth}{{l}*{6}{C}}
\Xhline{2\arrayrulewidth}
\addlinespace[0.1cm] 
 & \multicolumn{2}{c}{SimCLR} & \multicolumn{2}{c}{CPC} & \multicolumn{2}{c}{Multi-Task} \\ \cmidrule(lr){2-3} \cmidrule(lr){4-5} \cmidrule(lr){6-7}
\multirow{2}{*}{Metric} & \multirow{2}{*}{\shortstack{Replay\\SSL}} & \multirow{2}{*}{\shortstack{Fine-\\Tune}} & \multirow{2}{*}{\shortstack{Replay\\SSL}}  & \multirow{2}{*}{\shortstack{Fine-\\Tune}}  & \multirow{2}{*}{\shortstack{Replay\\SSL}}    & \multirow{2}{*}{\shortstack{Fine-\\Tune}} \\
& & & & & & \\
\Xhline{2\arrayrulewidth}
\addlinespace[0.1cm]
\rowcolor[gray]{0.9} \multicolumn{7}{l}{\textit{Galaxy S20 Ultra}} \\
Time (sec) & 9.66 & 16.31 & 73.54 & 21.46 & 6.06 & 25.18 \\
CPU (\%)   & 669.29 & 595.47 & 589.54 & 565.60 & 584.50 & 480.28 \\
Mem (MB)   & 83.38 & 89.29 & 339.90 & 36.86 & 99.84 & 48.41 \\

\rowcolor[gray]{0.9} \multicolumn{7}{l}{\textit{Raspberry Pi 4}} \\
Time (sec) & 33.62 & 21.64 & 330.29 & 54.87 & 20.04 & 40.37 \\
CPU (\%)   & 65.50 & 62.88 & 89.39 & 76.82 & 53.99 & 56.41 \\
Mem (MB)   & 62.31 & 71.01 & 714.63 & 712.62  & 47.45 & 57.53 \\

\rowcolor[gray]{0.9} \multicolumn{7}{l}{\textit{Raspberry Pi Zero 2 W (using flash memory for swap space)}} \\
Time (sec) & 88.45 & 46.43 & 4539.86 & 126.35 & 58.66 & 73.88 \\
CPU (\%)   & 61.32 & 66.57 & 32.48 & 79.83 & 57.78 & 61.49 \\
Mem (MB)   & 81.69 & 102.06 & 762.55 & 674.00 & 112.87 & 73.39 \\
\Xhline{2\arrayrulewidth}
\end{tabularx}}}
\vspace{-10pt}
\end{table}

\subsubsection{ReplaySSL Overhead}
\label{sec:computational_ovehread}

\shepherd{We aim to enable the practical deployment of pre-trained models to users, particularly those with resource-constrained mobile environments. To this end, we assess the computational feasibility of \proj{} for mobile devices, focusing on its user-side operations, ReplaySSL, and fine-tuning. We performed on-device training with three edge devices (Figure~\ref{fig:ondevice}): a Samsung Galaxy S20 Ultra, a Raspberry Pi 4, and a Raspberry Pi Zero 2 W. The Samsung Galaxy S20 Ultra had 12GB of RAM, and on-device training was implemented via Termux~\cite{termux}, a Linux terminal emulator for Android, to execute PyTorch-based training code. The Raspberry Pi 4 had 4GB of RAM and runs Ubuntu as its operating system. To simulate a more constrained environment, we used a Raspberry Pi Zero 2 W with only 512MB of RAM; lacking internal storage, we used flash memory as swap space.

Table~\ref{tab:overhead} presents the overhead of ReplaySSL and fine-tuning, measured independently across different self-supervised learning methods. On Samsung Galaxy S20 Ultra, all operations were completed in tens of seconds under the few-shot setting. Notably, ReplaySSL with SimCLR or multi-task learning took under 10 seconds, consuming under 100MB of memory—requiring even less time than fine-tuning. Consequently, all user-side operations of \proj{} could be performed in under 30 seconds on the smartphone. Although CPC required more resources, it was completed within 1.5 minutes while using approximately 340MB of memory. On Raspberry Pi 4, processing times increased by roughly three to four times compared to the smartphone; SimCLR and multi-task learning finished within one minute, and CPC took about six minutes, all while using under 1GB of memory. Even on the highly constrained Raspberry Pi Zero 2 W, SimCLR and multi-task learning remain feasible in under two minutes, whereas CPC's higher memory requirements trigger frequent swapping, prolonging execution to about 1.26 hours. Importantly, the adaptation step (ReplaySSL) is needed only once per user after obtaining the deployed model. Our findings confirm that end-users can conduct all necessary operations of \proj{} on the device with manageable computational overhead.}

\section{Discussion}

\subsection{Adapting to Changing Environments}

We designed \proj{} with a single domain adaptation step once self-supervised models are deployed to users. However, data characteristics within a single domain can change over time due to changing environments. This implies that the adapted model might not perform optimally as domain characteristics change continuously. We anticipate the potential for domain adaptation through ReplaySSL in such scenarios, as our adaptation step does not require user labels. This allows us to continuously adapt the model using the ongoing data stream from the user. Enhancing the efficiency and effectiveness of this approach in such dynamic scenarios is a direction for future work.

\subsection{Expanding Domain Coverage}

\shepherd{Our domain-specific task generation in MetaSSL is designed to reflect target-domain characteristics by composing tasks based on the same user or device information. In this approach, MetaSSL relies on sufficiently diverse domains during pre-training to learn effective adaptation strategies. If the pre-training data lack variety, the model may fail to generalize, highlighting the need for large-scale, diverse domain coverage.

As a future direction, more advanced meta-task generation approaches could be designed to handle the complexity of real-world applications. For instance, incorporating tasks involving a broader range of domains (\textit{e.g.}, different modalities and user contexts) would enable pre-trained models to adapt to varying deployment environments. Another promising approach involves reducing MetaSSL's reliance on domain diversity by devising methods that maintain generalizability with limited pre-training data, thus enhancing scalability. Ultimately, refining task generation to capture real-world variability will be a priority to ensure that MetaSSL remains robust across real-world applications.}

\subsection{Extending Self-Supervised Methods}
Our findings underscore the impact of domain shift on different self-supervised learning methods. We observed that the improvement from our domain adaptation varies with the type of self-supervised learning method applied. This suggests a need for a deeper understanding of how domain shifts affect various self-supervised learning approaches. Although our results shed light on the domain shift effects for established methods such as SimCLR, CPC, and multi-task learning, the behavior of numerous other self-supervised learning methods~\cite{haresamudram2020masked, xu2021limu, deldari2022cocoa, ouyang2022cosmo, jain2022collossl, eldele2023self} under domain shifts remains unexplored. Addressing this as future work is an essential step in the field.
\section{Conclusion}

\rev{We explored the domain shift challenge in mobile sensing, where self-supervised models are fine-tuned to heterogeneous domains. To address this, we proposed \proj{}, an adaptive meta-task replay approach for self-supervised learning. \proj{} combines MetaSSL, which uses meta-learning to produce self-supervised models prepared for domain adaptation, with ReplaySSL, an adaptation step that replays the meta-learned self-supervised task on target domain data. Our evaluation across different mobile sensing tasks demonstrates that \proj{} consistently outperforms existing self-supervised learning and domain generalization methods, achieving an average F1-score improvement of 9.4\%p. Additionally, \proj{} is computationally efficient, completing adaptation on a smartphone in under a few minutes. These findings validate \proj{} as a practical framework for enhancing pre-trained models for end-users with minimal overhead.}
\section*{Acknowledgments}

This work is funded in part by the National Research Foundation of Korea (NRF), funded by the Ministry of Science and ICT (MSIT) under grant RS-2024-00464269, the National Research Foundation of Korea (NRF) grant funded by the Korea government (MSIT) (RS-2024-00337007), and the Hong Kong GRF 16204224. 

\newpage

\balance

\bibliographystyle{ACM-Reference-Format}
\bibliography{main}


\end{document}